%% file: Analysis_THR_double.tex
\documentclass[10pt,twocolumn,english]{IEEEtran}
\usepackage[T1]{fontenc}
\usepackage{graphicx,color}
\usepackage{amsfonts}
\usepackage{amsmath, amssymb, amsthm}
\usepackage{algorithm}
\usepackage{algorithmic}
\usepackage{url}

\include{defs}

\usepackage{babel}
\makeatletter

\begin{document}

\title{Performance Guarantees of the Thresholding Algorithm for the Co-Sparse Analysis Model}

\author{Tomer Peleg, ~\IEEEmembership{Student Member,~IEEE}, and Michael Elad, ~\IEEEmembership{Fellow,~IEEE}
\thanks{T. Peleg is with the Department of Electrical Engineering, Technion -- Israel Institute of Technology, Haifa 32000, Israel (e-mail: tomerfa@tx.technion.ac.il). M. Elad is with the Computer Science Department, Technion -- Israel Institute of Technology, Haifa 32000, Israel (e-mail: elad@cs.technion.ac.il).}
\thanks{This work was supported by the European Commission's FP7-FET program, SMALL project (grant agreement no. 225913).}}

\makeatother
\maketitle

\begin{abstract}
\noindent The co-sparse analysis model for signals assumes that the signal of interest can be multiplied by an analysis dictionary $\mOm$, leading to a sparse outcome. This model stands as an interesting alternative to the more classical synthesis based sparse representation model. In this work we propose a theoretical study of the performance guarantee of the thresholding algorithm for the pursuit problem in the presence of noise. Our analysis reveals two significant properties of $\mOm$, which govern the pursuit performance: The first is the degree of linear dependencies between sets of rows in $\mOm$, depicted by the co-sparsity level. The second property, termed the Restricted Orthogonal Projection Property (ROPP), is the level of independence between such dependent sets and other rows in $\mOm$. We show how these dictionary properties are meaningful and useful, both in the theoretical bounds derived, and in a series of experiments that are shown to align well with the theoretical prediction. 
\end{abstract}


\section{Introduction}
\label{sec:Introduction}

Signal models lie at the core of various processing tasks, such as denoising, solving inverse problems, compression, interpolation, sampling, and more. One approach that has become very popular in the past decade is the synthesis-based sparse representation model. In this model, a signal $\vx\in \R^{d}$ is assumed to be composed as a linear combination of a {\em few} atoms (columns) from a dictionary $\mD\in \R^{d\times n}$ \cite{Review09,MyBook}. We typically consider a redundant dictionary with $n> d$.  The vector $\va\in \R^{n}$ is the sparse representation of the signal, i.e. $\|\va\|_0=k \ll d$. 

Vast work on the synthesis model during the past decade has been invested in an attempt to better understand it, and build practical tools for its use. The main activity concentrated on problems such as how to perform pursuit of the sparse representation from the possibly corrupted signal, deriving theoretical success guarantees for such pursuit algorithms, and techniques to learn the dictionary $\mD$ from signal examples. Referring specifically to the theoretical success guarantees, various measures were suggested along the years to formalize the notion of the suitability of a synthesis dictionary $\mD$ for sparse estimation. These include mutual coherence \cite{Donoho01,Donoho03}, the exact recovery condition (ERC) \cite{Tropp04}, the spark \cite{Donoho03} and the restricted isometry property (RIP) \cite{Candes05,Candes06}, the capacity sets \cite{Shtok08}, the characteristics for ``s-goodness'' \cite{Nemirovski11}, and others.

Using these measures, theoretical performance guarantees were developed for various synthesis pursuit algorithms in different setups. For example, the work presented in \cite{Schnass07} provided a coherence-based guarantee on the probability of success for the thresholding algorithm in a noise-free setup, under certain assumptions on the representation coefficients. A later work, \cite{Ben-Haim10}, suggested coherence-based performance guarantees for a wide range of pursuit algorithms, including the thresholding algorithm, in the presence of white Gaussian random noise. These two contributions are mentioned here since both these papers and the work reported here correspond to the simplest of all pursuit methods -- the thresholding algorithm.

While the {\em synthesis} model has been extensively studied, there is a dual {\em analysis} viewpoint to sparse representations that has only recently started to attract attention \cite{Elad07,Candes2011,Nam2011Sampta,Peyre2011Sampta,Nam2011ICASSP,Raja2011,Ophir2011,Davies2011,Nam2011Journal,PeyreAnalysis2011,
AnalysisKSVD11}. The analysis model relies on a linear operator (a matrix) $\mOm \in \R^{p\times d}$, which we will refer to as the {\em analysis dictionary}, and whose rows constitute {\em analysis atoms}. The key property of this model is our expectation that the analysis representation vector $\mOm \vx \in \R^{p}$ should be sparse with $\ell$ zeros. These zeros carve out the low-dimensional subspace that this signal belongs to. We shall assume that the dimension of this subspace, which is denoted by $r$ is indeed small, namely $r \ll d$.

While this description of the analysis model may seem similar to the synthesis counterpart approach, it is in-fact very different when dealing with a redundant dictionary $p>d$. Until recently, relatively little was known about this model, and little attention has been given to it in the literature, compared to the synthesis counterpart model. Several recent works have already started to treat some of the basic research questions arising from the analysis model, such as how to perform pursuit with this model \cite{Nam2011ICASSP,Nam2011Journal,AnalysisKSVD11}, what are the theoretical performance guarantees for the suggested pursuit algorithms \cite{Candes2011,Nam2011ICASSP,Raja2011,Nam2011Journal,PeyreAnalysis2011} and how to learn an analysis dictionary from a set of signal examples \cite{Peyre2011Sampta,Ophir2011,Davies2011,AnalysisKSVD11}. We shall return to some of these contributions towards the end of this paper, and discuss their relation to our work.

The main goal of this paper is a theoretical study of the analysis thresholding pursuit algorithm, deriving conditions for its success in recovering the co-support in the presence of additive noise. A by-product of this study is an identification of two complementary measures of goodness that characterize the analysis dictionary. The first is the degree of linear dependencies between rows in $\mOm$, which is depicted by the co-sparsity level. This property has already been noticed and discussed in previous works on the analysis model \cite{Nam2011Journal,AnalysisKSVD11}. The second property, termed the Restricted Orthogonal Projection Property (ROPP), is the level of independence between such dependent sets and other rows taken from the analysis dictionary. To the best of our knowledge, this is the first time that this property has been used in the published literature. In this paper we derive an explicit relation between these properties and the expected performance of analysis pursuit by means of thresholding. We demonstrate the goodness of our theoretical findings by matching them versus empirical performance results. 

This paper is organized as follows: In Section \ref{sec:AnalysisModel} we present the core concept of the analysis-based model, characterize the signals that belong to it, and discuss the notion of linear dependencies within the rows of the analysis dictionary. In Section \ref{sec:AnalysisTHR} we present the analysis pursuit problem of denoising a signal using the analysis model and suggest the thresholding algorithm for solving this problem. We test the performance of this algorithm in a series of synthetic experiments for different types of analysis dictionaries. A theoretical study of the performance of the analysis thresholding algorithm is conducted in Section \ref{sec:THRPerformance}. We begin by developing theoretical success guarantees for the thresholding algorithm and discuss the dictionary properties arising from this theoretical analysis. Then we revisit the empirical results in light of the developed theoretical guarantees. Section \ref{sec:Relations} discusses the relation of this work to existing contributions, and Section \ref{sec:Conclusion} concludes this paper.

\section{The Analysis Model and its Dictionary}

\label{sec:AnalysisModel}

\subsection{Basic Properties of the Analysis Model}

\label{subsec:AnalysisModelBasics}

This section begins with a brief review of the analysis-based model. The analysis model for the signal $\vx \in \R^d$ uses the possibly redundant analysis dictionary $\mOm \in \R^{p \times d}$, where redundancy here implies $p \ge d$. Throughout this paper the $j$th row in $\mOm$ will be denoted by $\vw_j^T$. A fundamental property of this model is the assumption that the analysis representation vector $\mOm \vx$ should be sparse. In this work we consider specifically $\ell_0$ sparsity, which implies that $\mOm \vx$ contains many zeros. The {\em co-sparsity} $\ell$ of the analysis model is defined as the number of zeros in the vector $\mOm \vx$,
\begin{equation}
\|\mOm \vx\|_0 = p-\ell.
\label{eq:DefEll}
\end{equation}

\noindent In this model we put an emphasis on the zeros of $\mOm \vx$, and define the {\em co-support} $\Lambda$ of $\vx$ as the set of $\ell = |\Lambda|$ rows that are orthogonal to it. In other words, $\mOm_\Lambda \vx =0$, where $\mOm_\Lambda$ is a submatrix of $\mOm$ that contains only the rows indexed in $\Lambda$. We also define the {\em co-rank} of a signal $\vx$ with co-support $\Lambda$ as the rank of $\mOm_\Lambda$. The signal $\vx$ is thus characterized by its co-support, which determines the subspace it is orthogonal to, and consequently the complement space to which it belongs. Just like in the synthesis model, we assume that the dimension of the subspace the signal belongs to, denoted by $r$, is small, namely $r \ll d$. The co-rank of such an analysis signal is $d-r$. How sparse can the analysis representation vector be? The answer to this question is directly related to the existence of linear dependencies within the rows of the analysis dictionary. This will become more clear in the next subsection where we discuss in detail the effect of having such dependencies on the possible co-sparsity levels.

\subsection{Linear Dependencies in the Analysis Dictionary}

\label{subsec:lin. dep.}

To motivate our discussion on the advantage of having linear dependencies within the rows of the analysis dictionary, let us first assume that the rows in $\mOm$ are in general-position, implying that every subset of $d$ or less rows are necessarily linearly independent. This is equivalent to the claim that the spark of $\mOm^T$ is full \cite{MyBook}. Naturally, for this case, $\ell < d$, since otherwise there would be $d$ independent rows orthogonal to $\vx$, implying $\vx = 0$. Thus, in this case the analysis model leads necessarily to a mild sparsity, $\|\mOm \vx\|_0 > p-d$, and for a highly redundant analysis operator, the cardinality of the analysis representation vector $\mOm \vx$ is expected to be quite high. In this case, the dimension of the subspace the signal belongs to is $r=d-\ell$. An example for such a dictionary is a Gaussian random one, denoted $\mOm_{RAND}$, where the rows are drawn identically and independently from a normal distribution.

A more interesting case is when $\mOm^T$ has \emph{non-full spark}, implying that linear dependencies exist between the dictionary atoms. The immediate implication is that $\ell$ {\em could go beyond} $d$, and yet the signal would not necessarily be nulled. An example of such a dictionary is the set of cyclic horizontal and vertical one-sided derivatives, applied on a 2D signal of size $\sqrt{d}\times \sqrt{d}$. The corresponding analysis dictionary, denoted $\mOm_{DIF}$, is of size $2d\times d$, thus twice redundant. This dictionary was discussed in detail in \cite{Nam2011Journal}, showing that its rows exhibit strong linear dependencies.

Note that if we perform right multiplication of an analysis dictionary $\mB$ by an invertible square matrix $\mA$ then the resulting analysis dictionary $\mOm\doteq\mB\mA$ exhibits the same linear dependencies between its rows as in $\mB$. To see that this is indeed true, let $\Lambda\subseteq\{1,\ldots,p\}$ and suppose that there exists a vector $\vg\in\R^{\ell}$ such that $\vg^T\mB_{\Lambda}=0$, namely the rows of $\mB_{\Lambda}$ are linearly dependent. Then $\vg$ also satisfies $\vg^T\mOm_{\Lambda}=\vg^T\mB_{\Lambda}\mA=0$. For example, the rows of the analysis dictionary that is generated as $\mOm_{MIX}=\mOm_{DIF}\mA$, where $\mA$ is a square matrix consisting of $d$ Gaussian random rows, exhibit the same linear dependencies as $\mOm_{DIF}$.

\begin{figure}
\centering
\includegraphics[scale=0.4]{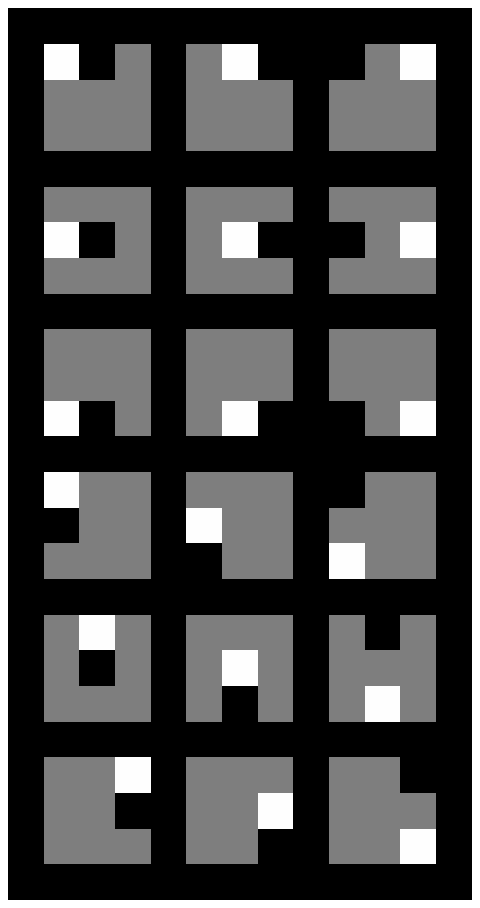}
\includegraphics[scale=0.4]{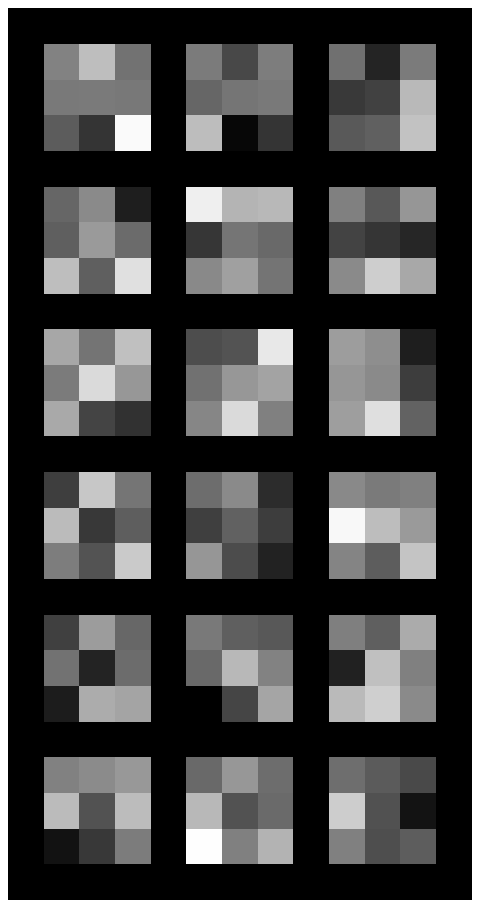}
\includegraphics[scale=0.4]{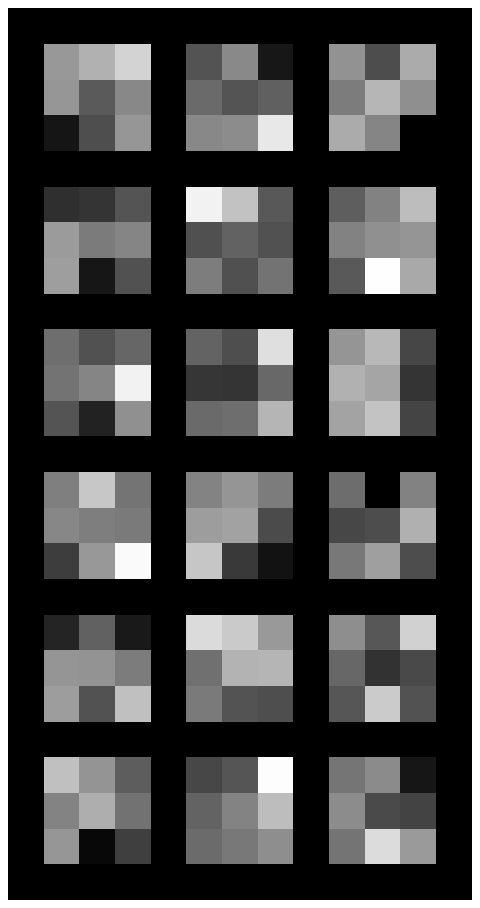}
\caption[]{Three types of analysis dictionaries of size $18\times 9$: Left - $\mOm_{DIF}$, Middle - $\mOm_{RAND}$, Right - $\mOm_{MIX}$. Each dictionary atom is displayed as a $2$D patch of size $3$-by-$3$.}
\label{fig:OmegaTypes}
\end{figure}

Fig. \ref{fig:OmegaTypes} shows the three types of dictionaries mentioned above for $p=18,~d=9$. Throughout this paper we will experiment with these three dictionaries. The reason for such low dimensional matrices is the fact that the study of the properties of the analysis dictionary will require exhaustive computations over all possible $2^p$ co-supports. In particular, these dictionary properties will appear in the performance guarantees we are about to derive for the analysis thresholding algorithm (see Section \ref{subsec:THRGuarantees}). Towards the end of this paper we will replace the exact dictionary properties by approximate ones, which are obtained from a set of signal examples generated from the dictionary. This will allow us to show theoretical results also for higher dimensions and check how well they predict the empirical results (see the end of Section \ref{subsec:THRPerformanceRevisited}).

\begin{figure}
\centering
\includegraphics[scale=0.4]{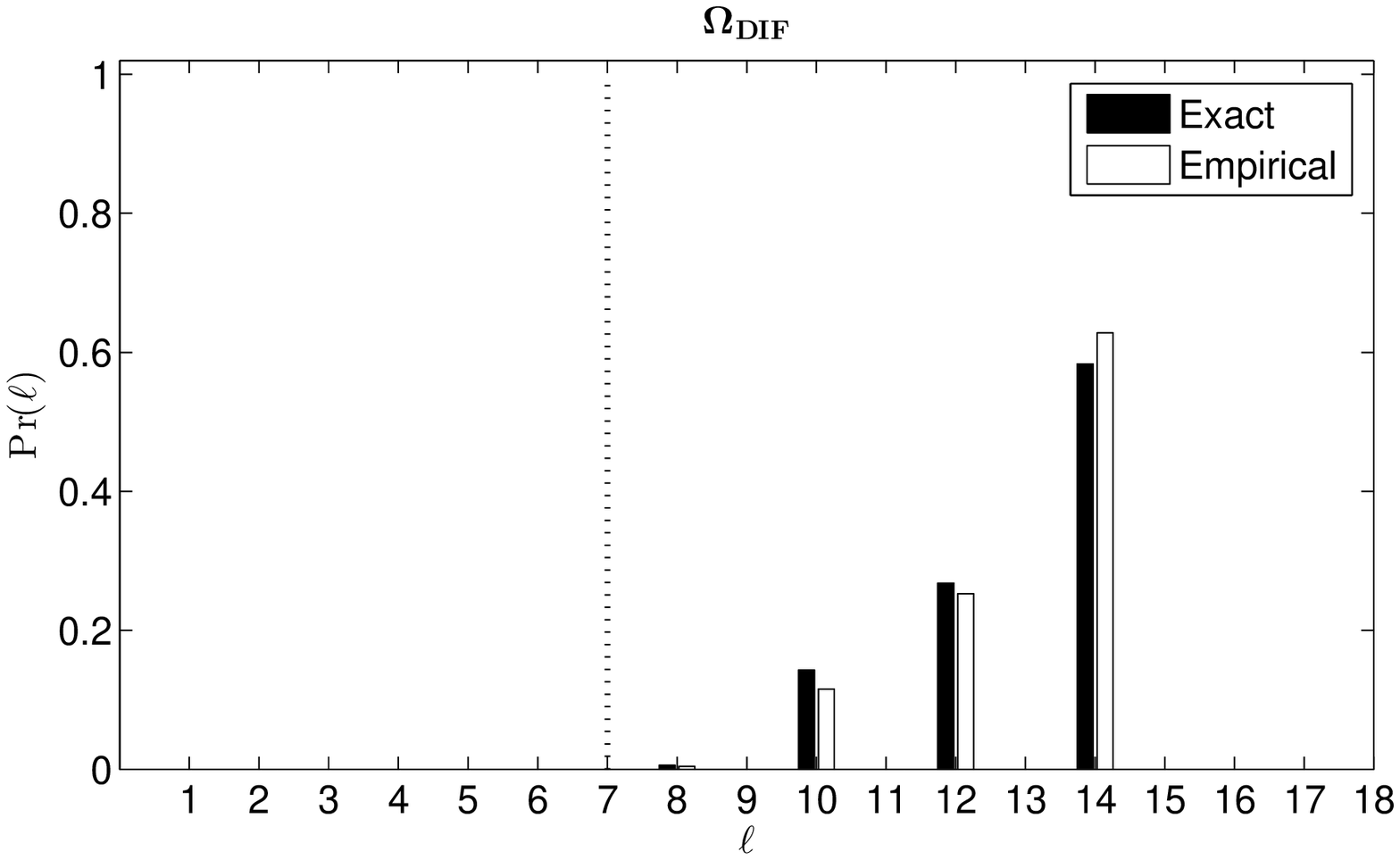}\\
\vspace{10pt}
\includegraphics[scale=0.4]{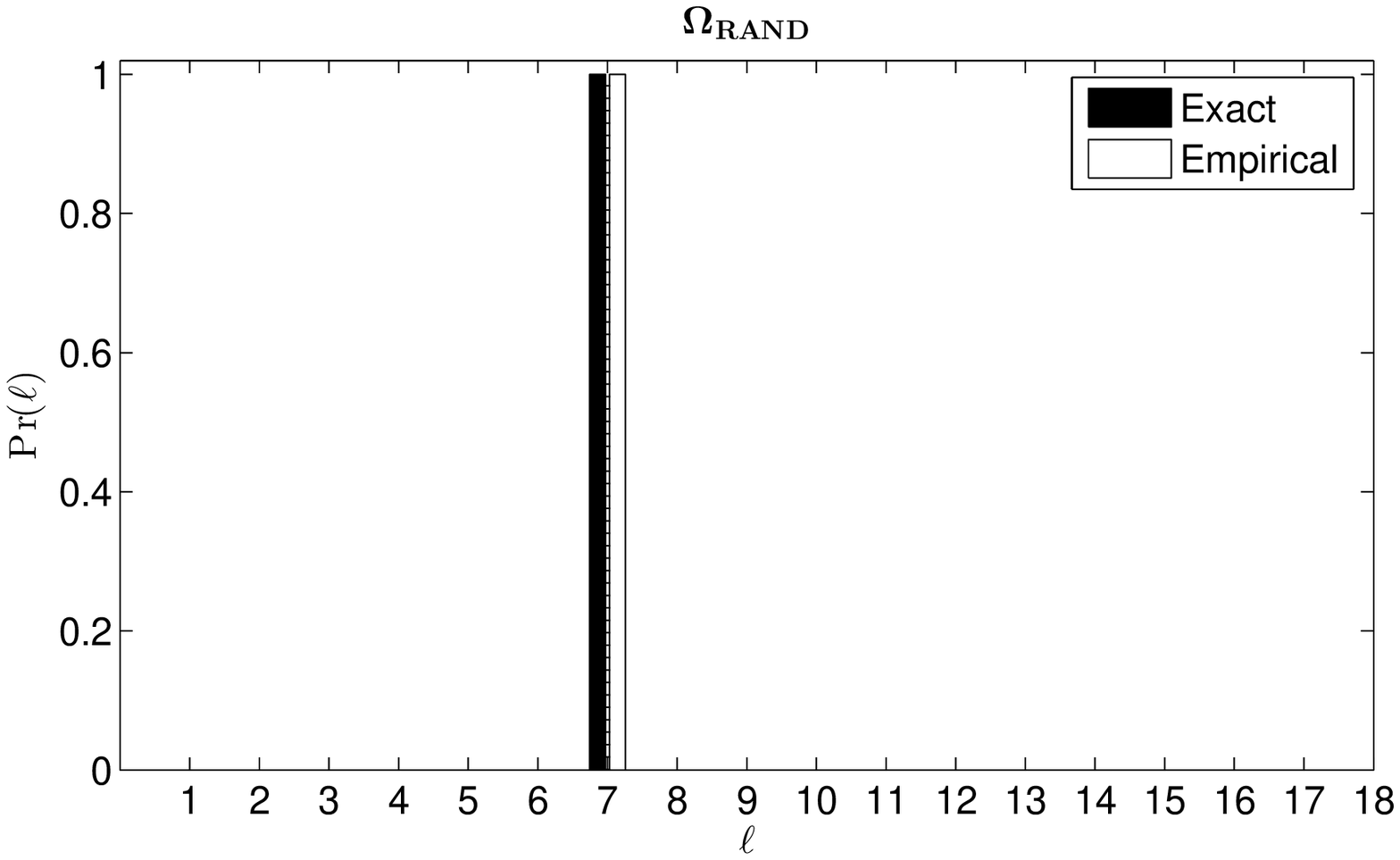}\\
\vspace{10pt}
\includegraphics[scale=0.4]{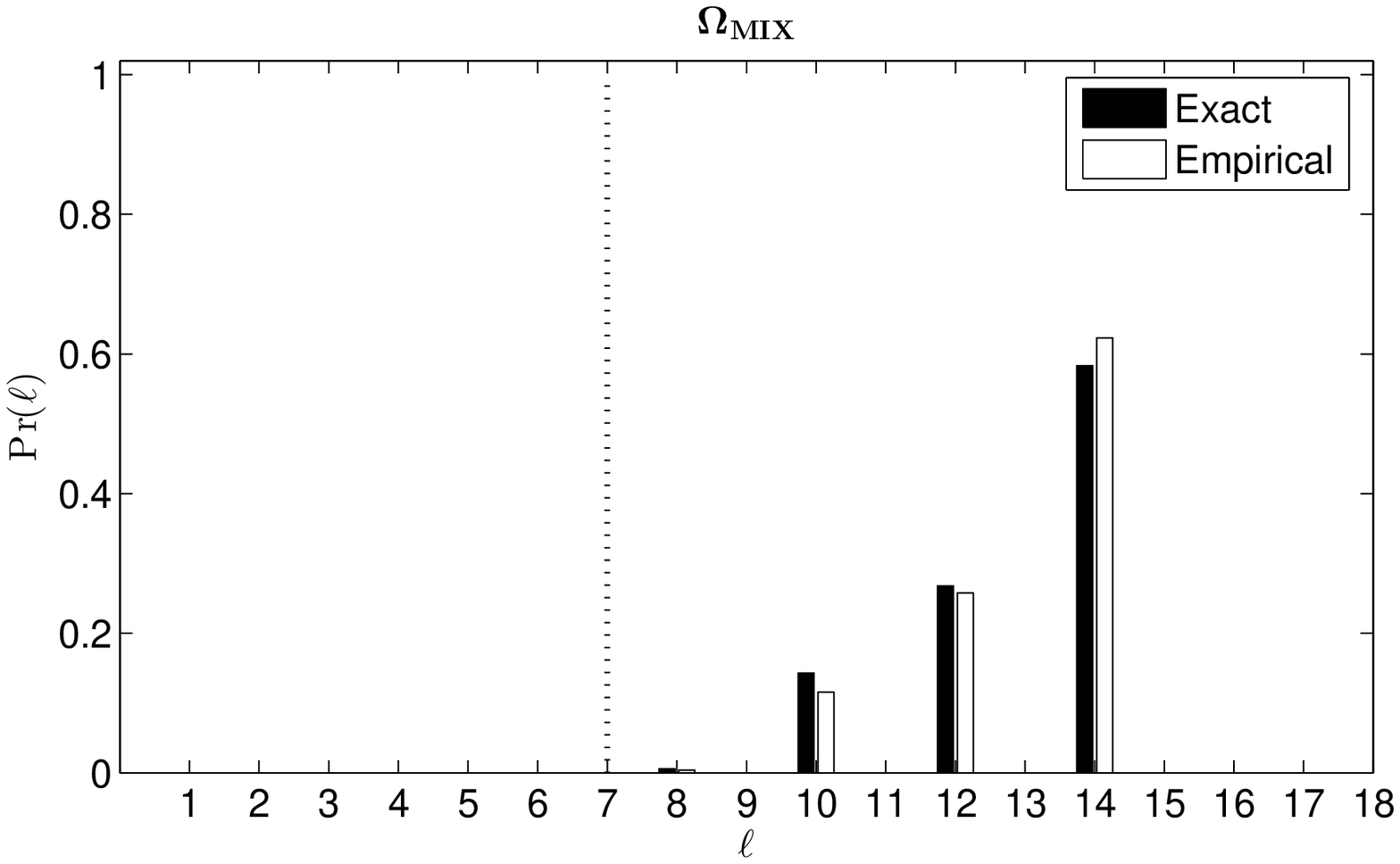}
\caption[]{The effective co-sparsities corresponding to each type of analysis dictionary of size $18\times 9$: Top - $\mOm_{DIF}$, Middle - $\mOm_{RAND}$, Bottom - $\mOm_{MIX}$. For each type we show the exact co-sparsity distribution, which is computed exhaustively for all possible co-supports corresponding to a co-rank of $7$. We also show an empirical normalized histogram, which is computed from $10,000$ analysis signals of co-rank $7$ that were generated using the process described in the beginning of Section \ref{subsec:THRExperiments}. The reference value of $\ell=7$ is indicated by the vertical dotted line. As can be seen, the effective co-sparsities are all strictly higher for both $\mOm_{DIF}$ and $\mOm_{MIX}$.}
\label{fig:OmegaCoSparsity}
\end{figure}

As mentioned above, when the rows in $\mOm$ are not in general-position, the co-sparsity $\ell$ can be greater than $d$. This behavior is demonstrated in Fig. \ref{fig:OmegaCoSparsity} showing the distributions of $\ell$ for the three types of $\mOm$ shown in Fig. \ref{fig:OmegaTypes} and co-rank $7$. For each type the exact co-sparsity distribution is computed exhaustively for all possible co-supports corresponding to a co-rank of $7$. We also show an empirical normalized histogram, which is computed from $10,000$ analysis signals of co-rank $7$ that are generated using the process that will be described in the beginning of Section \ref{subsec:THRExperiments}. As can be seen the distribution for $\mOm_{DIF}$ and $\mOm_{MIX}$ coincide, as should be expected from the observation mentioned above (both dictionaries exhibit the same linear dependencies between their rows). In both cases, though the signals have a fixed co-rank $7$, their actual co-sparsities are much higher, varying in the range $8$ to $14$. Interestingly, odd co-sparsity values cannot lead to the chosen co-rank, as indeed seen in Fig. \ref{fig:OmegaCoSparsity}. Thus, we see that by allowing linear dependencies between the rows in $\mOm$, co-sparsities much higher than the signal dimension $d$ can be achieved.

\begin{figure}
\centering
\includegraphics[scale=0.4]{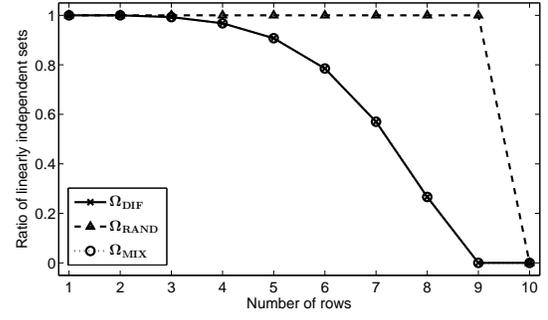}
\caption[]{The signatures for three types of analysis dictionary of size $18\times 9$ that were shown in Fig. \ref{fig:OmegaTypes}. As can be seen, both $\mOm_{DIF}$ and $\mOm_{MIX}$ have the same signature, which is strictly lower than $1$ for $k\geq 3$. Therefore the spark of these dictionaries is $3$, namely it is non-full. For $\mOm_{RAND}$ however the signature equals $1$ for all $k=1,\ldots,9$ and therefore its spark is $d+1=10$.}
\label{fig:OmegaSignature}
\end{figure}

An alternative measure for the linear dependencies between sets of rows in $\mOm$ is the signature of the analysis dictionary, which is defined as the ratio of linearly independent sets of $k$ rows out of all possible sets of size $k$ -- this ratio is denoted by $f(k)$ \cite{Elad06}. Since every set of size at least $d+1$ is necessarily linearly dependent, it is sufficient to compute the ratios mentioned above for $k=1,\ldots,d$. The spark of $\mOm^T$ can be readily computed from the signature $f(k)$ -- it is the smallest index $k$ such that $f(k)<1$. The signatures of the three analysis dictionaries that were shown in Fig. \ref{fig:OmegaTypes} are depicted in Fig. \ref{fig:OmegaSignature}. Clearly, $\mOm_{DIF}$ and $\mOm_{MIX}$ have the same signature, as they exhibit the same linear dependencies. Their signature is much lower than for $\mOm_{RAND}$ whose signature equals $1$ for all $k=1,\ldots,d$. We observe that the spark of $\mOm_{DIF}^T$ and $\mOm_{MIX}^T$ is $3$, whereas the spark of $\mOm_{RAND}^T$ is $d+1=10$ (i.e. the spark is full). To conclude this section, note that a lower dictionary signature indicates that there are more linear dependencies within its rows, and these allow for larger co-sparsity levels.

\section{Analysis Thresholding}

\label{sec:AnalysisTHR}

\subsection{Analysis Pursuit}

\label{subsec:AnalysisPursuit}

In this paper we assume that $\vx$ is a co-sparse analysis signal with co-rank $d-r$, and this signal is contaminated by additive noise, $\vy = \vx + \ve$. Starting with the {\em oracle} setup, where the true co-support $\Lambda$ is known, we can simply recover $\vx$ by projecting $\vy$ onto the subspace orthogonal to $\mOm_{\Lambda}$:
\begin{equation}
\hat{\vx}=\left(\mI-\mOm_{\Lambda}^\dagger\mOm_{\Lambda}\right)\vy.
\label{eq:oracle formula}
\end{equation}
Assuming a deterministic signal $\vx$ residing in a $r$-dimensional analysis subspace and white and zero-mean Gaussian noise $\vv$ with variance $\sigma^2$, the mean denoising error in the oracle setup is given by
\begin{equation}
\E\|\vx - \hat{\vx}\|_2^2=\tr\left(\mI - \mOm_{\Lambda}^\dagger
\mOm_{\Lambda}\right)\sigma^2=r\sigma^2,
\label{eq:oracle error}
\end{equation}
where $\tr(\cdot)$ denotes the trace of a matrix. For more details see \cite{AnalysisKSVD11}.

In the general case the correct co-support is unknown and it should be estimated from $\vy$. Recovering the noise-free signal $\vx$ requires solving a problem of the form
\begin{align}
\left\{{\hat \vx},~{\hat \Lambda} \right\} =
\argminlong{\vx,\,\Lambda} ~ \|\vx - \vy\|_2 \quad &\st\label{eq:analysis-rank1}\\
&\mOm_\Lambda\vx=0 \nonumber \\
&\rank\left(\mOm_\Lambda\right)=d-r \nonumber
\end{align}
\noindent We refer to this problem as the analysis sparse-coding or analysis-pursuit. This problem can be readily reformulated as a two-step recovery process. To eliminate the dependency on $\vx$ we can place the oracle formula of (\ref{eq:oracle formula}) into the problem of (\ref{eq:analysis-rank1}). We get that recovering the co-support $\Lambda$ results in solving the problem
\begin{align}
\hat \Lambda = \argminlong{\Lambda} ~ \|\mOm_{\Lambda}^\dagger\mOm_{\Lambda}\vy\|_2 \quad &\st\label{eq:analysis-rank2}\\
\rank\left(\mOm_\Lambda\right)=d-r \nonumber
\end{align}
\noindent Once the co-support has been recovered we can project $\vy$ onto the orthogonal subspace (using (\ref{eq:oracle formula})), just as in the oracle setup.

Similar to the synthesis sparse approximation problem, the problem posed in Eq. (\ref{eq:analysis-rank1}) is combinatorial in nature and can thus only be approximated in general. One approach for approximating the solution is to use a relaxed $\ell_1$ penalty function on the coefficients $\mOm\vx$, producing
\begin{equation}
\hat\vx = \argminlong{\vx}~ \|\vx - \vy\|_2 \quad\st\quad
\|\mOm\vx\|_1\le T.
\label{eq:l1analysis-sparsity}
\end{equation}
\noindent This approach is parallel to the basis-pursuit approach for synthesis approximation \cite{BP98}. A second approach parallels the synthesis greedy pursuit algorithms \cite{MP93,OMP93} and suggests selecting rows from $\mOm$ one-by-one in a greedy fashion. The solution can be built by either detecting the rows that correspond to the non-zeros in $\mOm\vx$, or by detecting the zeros. The GAP algorithm, described in \cite{Nam2011Journal}, aims at detecting the non-zeros, whereas the BG and OBG algorithms developed in \cite{AnalysisKSVD11} detect the zeros. 

\subsection{The Thresholding Algorithm}

\label{subsec:THRAlgorithm}

In this work we will take the alternative (and simpler) approach of thresholding. This algorithm computes the analysis representation $\mOm\vy$ and chooses the smallest entries as the estimated co-support. Thresholding will always obtain a perfect recovery of the co-support in noise-free setups since $\mOm_\Lambda\vx=0$ and $|\vw_j^T\vx|>0$ for all $j\in\Lambda^C$. We suggest using it also in the presence of noise. A detailed description of the analysis thresholding algorithm is given in Algorithm \ref{alg:THR}.

\begin{algorithm}[htbp]
\caption{\sc ~~Analysis Thresholding Algorithm} \label{alg:THR}
\begin{algorithmic}[1]
\bigskip
\STATE {\bf Input:} Analysis dictionary $\mOm\in\R^{p\times d}$,
signal $\vy\in\R^d$, and target co-rank $d-r$

\STATE {\bf Output:} Signal ${\hat \vx}\in\R^d$ with co-rank $d-r$ approximating the minimization of $\|\vy-{\hat\vx}\|_2$ and its co-support $\hat \Lambda$

\medskip
\STATE {\bf Inner Products:} $z_k :=\,|\,\vw_k^T\,\vy\,|,\forall k=1,\ldots,p$
\STATE {\bf Sort:} Set $\Gamma$ to be the index set $\{\,1,\ldots,p\,\}$ sorted by the value of $z_k$ in increasing order
\STATE {\bf Initialization:} Set $i=0$, $\Lambda := \emptyset$

\smallskip
\WHILE{$\rank\left(\mOm_{\Lambda}\right)<d-r$}

\STATE $i :=i+1$
\STATE {\bf Update Co-Support:} $\Lambda := \Lambda \,\cup\,\{\,\Gamma_i\,\}$

\ENDWHILE

\smallskip
\STATE {\bf Project:} $\hat \vx = \left(\mI - \mOm_\Lambda^\dagger \mOm_\Lambda\right)\vy$
\STATE {\bf Refine Co-Support} $\hat \Lambda =\{k\,|\,1 \le k \le p,\,|\,\vw_k^T\,\hat \vx\,|\,<\epsilon_0\}$
\bigskip
\end{algorithmic}
\end{algorithm}

The process begins by computing the inner products between all the rows in $\mOm$ and the signal $\vy$ and sorting the index set $\{\,1,\ldots,p\,\}$ according to the magnitudes of these inner products in increasing order, resulting in a new index set $\Gamma$. The co-support is initialized to be an empty set. We then accumulate rows into the co-support, in a row-by-row fashion, according to their order of appearance in the set $\Gamma$. This process repeats until the target co-rank is achieved, namely $\rank\left(\mOm_{\Lambda}\right)=d-r$. The solution $\hat \vx$ is then computed by projecting $\vy$ onto the subspace orthogonal to the selected rows. Finally, the co-support is refined by recalculating the representation vector $\mOm \hat \vx$ and finding the additional coefficients that fall below some small threshold $\epsilon_0$. This can reveal additional rows that are orthogonal to the signal estimate, namely the rows that are spanned by the existing set of rows $\mOm_{\Lambda}$. Despite the fact that the last step (``Refine Co-Support``) has no impact on the signal recovery, it is still significant for our purposes, as our study checks the correctness of the found co-support.

In practice, the above algorithm can be implemented efficiently by accumulating an orthogonalized set of the co-support rows using a modified Gram-Schmidt process. This process is applied according to the order of appearance in the set $\Gamma$. Denoting by $\{\vq_j\}_{j=1}^{J}$ the orthogonal set accumulated so far (as column vectors), the orthogonalization of a new row $\vw_{\Gamma_i}^T$ is obtained by
\begin{equation}
\vq_i = \vw_{\Gamma_i} - \sum_{j=1}^{J} (\vq_j^T \vw_{\Gamma_i})  \vq_j. \label{eq:orthogonalization}
\end{equation}
If $\vq_i$ equals zero, it is not added to the orthogonal set, as it is already spanned by the existing one. Otherwise, this vector is normalized, $\vq_i = \vq_i/\|\vq_i\|_2$.

The above-described orthogonalization process allows us first of all to avoid the computation of the rank of the submatrix $\mOm_{\Lambda}$, since the number of vectors in the orthogonalized set ($J$) equals the desired rank. Secondly, the orthogonalized set $\{\vq_j\}_{j=1}^{d-r}$ can also be used to avoid the matrix inversion in the ``Projection`` step, which translates comfortably to
\begin{eqnarray}
{\hat \vx}_i  = \left(\mI-\mOm_{\Lambda_i}^\dagger\mOm_{\Lambda_i}\right) \vy = \left[\mI -\sum_{j=1}^i \vq_j\vq_j^T \right]\vy.
\label{eq:ProjStep}
\end{eqnarray}

\subsection{Synthetic Experiments}

\label{subsec:THRExperiments}

We now demonstrate how the thresholding algorithm (see Algorithm \ref{alg:THR}) performs through a series of synthetic experiments. Throughout this subsection we shall assume that the analysis signals are generated by the following process: Choose randomly a set of row indices $\Lambda\subseteq\{1,\ldots,p\}$, which will be the signal's co-support. Starting with a random vector $\vu$, whose entries are assumed to be drawn independently and identically from a zero-mean Gaussian distribution with variance $\sigma_u^2$, project it onto the subspace orthogonal to $\mOm_\Lambda$:
\begin{equation}
\vx = (\mI-\mOm_\Lambda^\dagger\mOm_\Lambda)\vu,
\label{eq:analysis-projection}
\end{equation}
\noindent and $\vx$ is an analysis signal that satisfies our co-sparsity assumption. For a general-positioned $\mOm$ we choose exactly $\ell$ rows from $\mOm$ at random. Otherwise we choose $d-r$ linearly independent rows from $\mOm$. Once a signal $\vx$ has been generated, its analysis representation $\mOm\vx$ is re-computed, possibly revealing additional rows that are orthogonal to this signal, due to linear dependence on the chosen subset $\Lambda$.

\begin{figure*}
\centering
\includegraphics[scale=0.5]{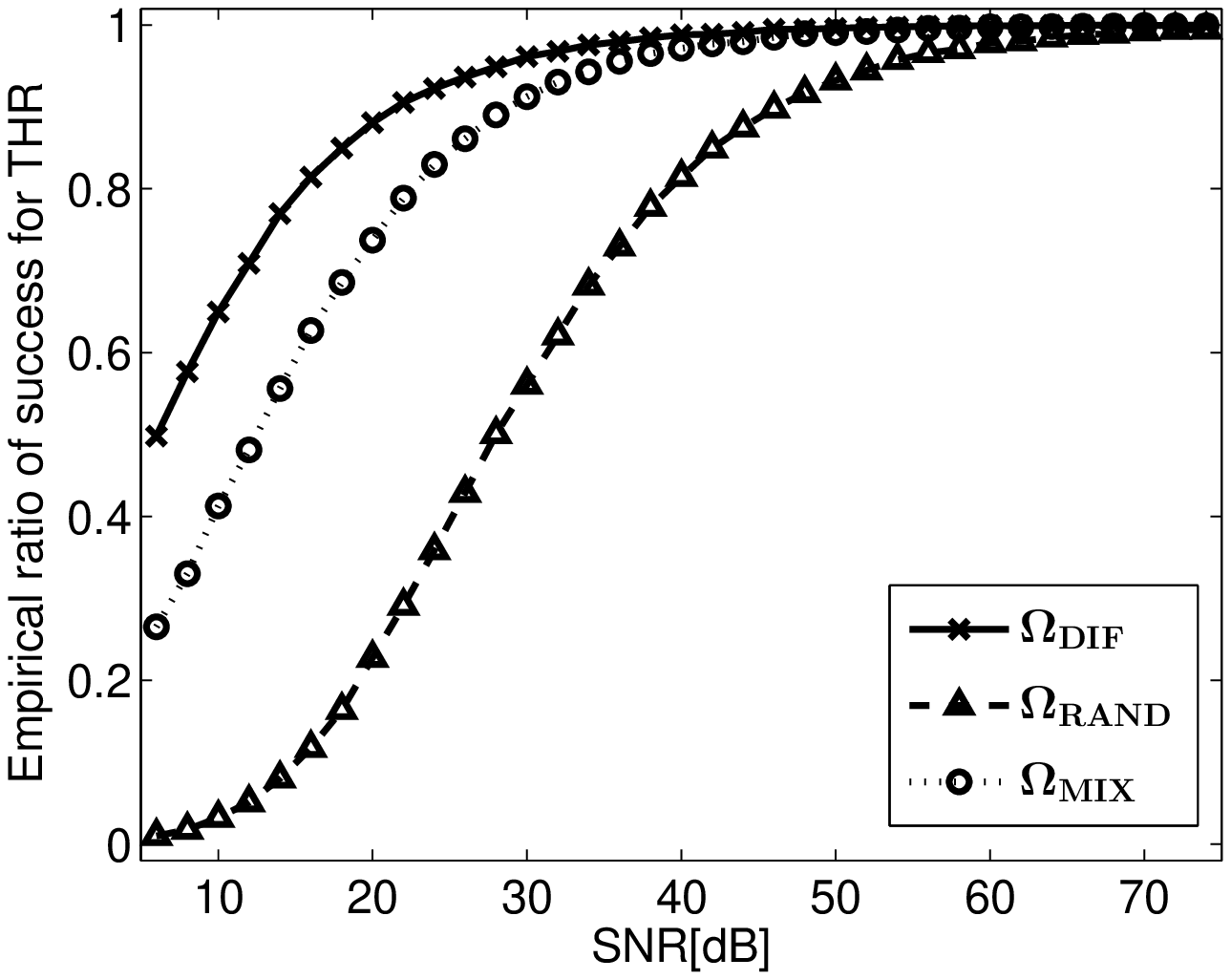}
\hspace{10pt}
\includegraphics[scale=0.5]{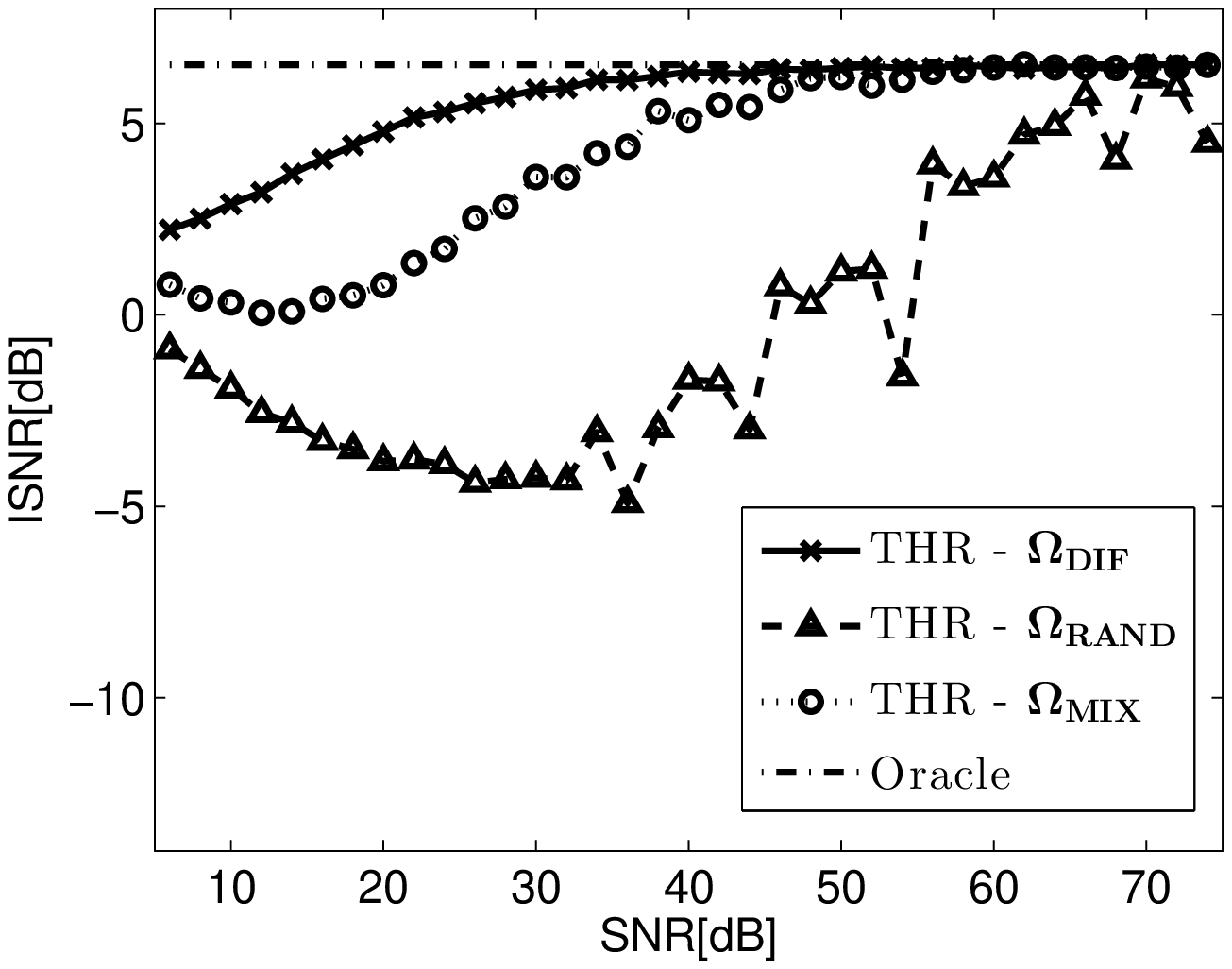}
\caption[]{Denoising experiments with analysis signals of co-rank $7$ created from the three types of analysis dictionaries of size $18\times 9$ that were shown in Fig. \ref{fig:OmegaTypes}. Additive white noise is added to each of these signals for varying noise levels and then the thresholding algorithm (see Algorithm \ref{alg:THR}) is applied on each signal to obtain a recovery of its co-support and its resulting denoised signal. Top: The empirical probability of success in recovering the true co-support for the thresholding algorithm on each of the dictionary types. Bottom: The noise attenuation performance obtained for the thresholding algorithm on each of the dictionary types. These are compared with the oracle result, where denoising is obtained by projection onto the correct analysis subspace (knowing the true co-support of the signals).}
\label{fig:EmpiricalResultsTHR}
\end{figure*}

We generate $N=10,000$ analysis signals in $\R^9$ residing in $2$-dimensional subspaces for the three types of analysis dictionaries shown in Fig. \ref{fig:OmegaTypes} -- normalized histograms of their effective co-supports are depicted in Fig. \ref{fig:OmegaCoSparsity}. These signals are contaminated with additive white Gaussian noise at different noise levels $\sigma$, resulting in a set of noisy signals $\left\{y_j\right\}_{j=1}^N$ for each dictionary type and noise level. The thresholding algorithm is then applied on these signals with a target co-rank of $d-r=7$. Results are shown in Fig. \ref{fig:EmpiricalResultsTHR} for various signal-to-noise ratios (SNR) in the range $6dB$ to $74dB$. Each SNR level is related to the ratio $\sigma/\sigma_u$ by
\begin{equation}
SNR\doteq10\log_{10}\left(\frac{\E\|\vx\|_2^2}{\E\|\vy-\vx\|_2^2}\right)=-20\log_{10}\left(\sqrt{\frac{d}{r}}\frac{\sigma}{\sigma_u}\right).
\label{eq:SNR computation}
\end{equation}
\noindent where in the last equation we used the equation $\E\|\vx\|_2^2=\tr\left(\mI - \mOm_{\Lambda}^\dagger\mOm_{\Lambda}\right)\sigma_u^2=r\sigma_u^2$, which holds since $\vx$ is a zero-mean Gaussian vector with a covariance matrix $\left(\mI - \mOm_{\Lambda}^\dagger\mOm_{\Lambda}\right)\sigma_u^2$ (exhibiting a similar form as in the oracle error -- see Eq. (\ref{eq:oracle error})), and $\E\|\vy-\vx\|_2^2=d\sigma^2$. At this point we should mention that the SNR levels shown on the right part of the figure are very high ones (for example SNR=$60dB$ means that the signal energy is $1000$ times the noise energy). Setups with such high SNR levels can be considered as almost noise-free. Therefore we expect that the thresholding algorithm will obtain a perfect recovery of the co-support in these setups, just like in the noise-free setup.

In Fig. \ref{fig:EmpiricalResultsTHR} we can see on the top the empirical probability of success for the thresholding algorithm on each of the dictionaries. Note that ``success'' refers here to an exact recovery of the true co-support. On the bottom we can see the denoising performance, measured as the average SNR improvements (ISNR):
\begin{equation}
ISNR\doteq-10\log_{10}\left(\frac{\|\hat\vx-\vx\|_2^2}{d\sigma^2}\right)
\label{eq:ISNR}
\end{equation}
\noindent These are also compared with the oracle performance, which corresponds to an ISNR of $-10\log_{10}\left(r/d\right)=6.53dB$. We can see at the top right corner of the figure that thresholding succeeds with probability one for all three types of dictionaries, which aligns with our expectations for high SNRs that were mentioned before.

Several important observations can be drawn from the results shown in Fig. \ref{fig:EmpiricalResultsTHR}. First of all, we can see that the probability of success decreases as the SNR deteriorates. This aligns with the simple intuition that the higher the noise, the higher the chance of any pursuit algorithm to make mistakes in the co-support detection. Second, the highest success ratio and ISNR are obtained for $\mOm_{DIF}$ at all noise levels; the second-best results relate to $\mOm_{MIX}$ and the worse to $\mOm_{RAND}$.

The observation that $\mOm_{RAND}$ exhibits the worst performance does not come as a surprise to us. The fact that having many linear dependencies in an analysis dictionary $\mOm$ leads to better denoising results has already been observed in a previous work \cite{AnalysisKSVD11}. However, the performance gap between $\mOm_{DIF}$ and $\mOm_{MIX}$ is not obvious at all, if we recall that both exhibit the same linear dependencies between their rows (and hence the same co-sparsity distribution). This calls for a deeper theoretical study of the thresholding algorithm, which is the topic of the next section.

\section{Theoretical Study of Analysis Thresholding}

\label{sec:THRPerformance}

This section consists of the main contribution of this paper: A theoretical analysis of the capability of the thresholding algorithm to recover the true analysis co-support in the presence of additive noise, and the implications of this analysis. We start in Section \ref{subsec:THRGuarantees} with the derivation of our main result -- a lower-bound on the probability of successfully recovering the co-support by the analysis thresholding algorithm. Section \ref{subsec:Discussion} discusses the obtained results and specifically the meaning of the measures proposed for the analysis dictionary. In Section \ref{subsec:THRPerformanceRevisited} we revisit these results in an attempt to explain them further, and contrast them with the empirical evidence we have just created. As this work focuses on the probability of the analysis thresholding algorithm to recover the exact co-support, the relative denoising performance will not be further explored in this paper and remains a topic for future research.

\subsection{Theoretical Guarantees for Analysis Thresholding}

\label{subsec:THRGuarantees}

Before we turn to the development of the theoretical guarantees for the analysis thresholding algorithm, we would like to set some basic assumptions and notations.
First, we assume that all the rows in $\mOm$ have unit-norm. Secondly, we denote an index set of $d-r$ linearly independent rows taken from $\Lambda$ by $\widetilde{\Lambda}\subseteq\Lambda$, namely $\Span\left\{\mOm_{\widetilde{\Lambda}}\right\}=\Span\left\{\mOm_\Lambda\right\}$. Finally, given a noise-free signal $\vx$ and an analysis dictionary $\mOm$, let us define 
\begin{equation}
\zmin\doteq \min{j\in \Lambda^C}{\left|\vw_j^T\vx\right|}, 
\label{eq:zmin}
\end{equation}
\noindent where $\Lambda$ is the co-support of $\mOm\vx$ and $\Lambda^C$ is the complementary index set. For the co-sparse analysis signal $\vx$ we have that $\mOm_\Lambda\vx = 0$, implying that $\mOm_{\Lambda^C}\vx \neq 0$. The value of $\zmin$ is the smallest of those non-zero inner-products with $\Omega_{\Lambda^C}$, and it plays a major role in the ability of the thresholding algorithm to tell the right co-support rows from the rest in the noisy case. We begin our performance study of this algorithm with a sufficient condition on $\zmin$ for success.

\begin{lemma}
Let $\vy=\vx+\ve$, where $\vx$ is a co-sparse analysis signal with co-support $\Lambda$ on $\mOm$. If $\vx$ and $\mOm$ satisfy 
$\zmin\geq 2~\text{Max}_{j\in\widetilde{\Lambda}\cup\Lambda^C}~\left|\vw_j^T\ve\right|$, then the thresholding algorithm succeeds in recovering the true co-support $\Lambda$ of $\vx$ from $\vy$.
\label{lemma:THRGeneralCond}
\end{lemma}

\begin{IEEEproof}
We begin with the simple observation that the thresholding algorithm succeeds in recovering the true co-support $\Lambda$ of $\vx$ when
\begin{equation}
\max{j\in\widetilde{\Lambda}}~\left|\vw_j^T\vy\right|<\min{j\in\Lambda^C}~\left|\vw_j^T\vy\right|.
\label{eq:CondInnerProd}
\end{equation}

\noindent Since $\vw_j^T\vx=0$ for all $j\in\widetilde{\Lambda}$ the left-hand side of (\ref{eq:CondInnerProd}) translates to
\begin{equation}
\max{j\in\widetilde{\Lambda}}~\left|\vw_j^T\vy\right|=\max{j\in\widetilde{\Lambda}}~\left|\vw_j^T\ve\right|.
\label{eq:InnerProdLeft}
\end{equation}
\noindent For the right-hand side of (\ref{eq:CondInnerProd}) we derive a lower bound
\begin{equation}
\min{j\in\Lambda^C}~\left|\vw_j^T\vy\right|\geq\min{j\in\Lambda^C}~\left|\vw_j^T\vx\right|-\left|\vw_j^T\ve\right|
\geq\zmin-\max{j\in\Lambda^C}~\left|\vw_j^T\ve\right|,
\label{eq:InnerProdRight}
\end{equation}
\noindent where the first inequality holds from the triangle inequality and the second holds from the properties of the minimum and maximum operators,
\begin{equation}
\text{Min}~(f-g)\geq\text{Min}~f+\text{Min}~(-g)=\text{Min}~f-\text{Max}~g.
\label{eq:MinMax}
\end{equation}

\noindent From (\ref{eq:CondInnerProd})-(\ref{eq:InnerProdRight}) we get that a sufficient condition for success of the thresholding algorithm is:
\begin{equation}
\max{j\in\widetilde{\Lambda}}~\left|\vw_j^T\ve\right|<\zmin-\max{j\in\Lambda^C}~\left|\vw_j^T\ve\right|,
\label{eq:ZminCondInter1}
\end{equation}
\noindent which can be comfortably replaced by the sufficient condition
\begin{equation}
\zmin>2\max{j\in\widetilde{\Lambda}\cup\Lambda^C}~\left|\vw_j^T\ve\right|,
\label{eq:ZminCond}
\end{equation}
\noindent since 
\begin{equation}
2\max{j\in\widetilde{\Lambda}\cup\Lambda^C}~\left|\vw_j^T\ve\right|\geq\max{j\in\widetilde{\Lambda}}~\left|\vw_j^T\ve\right|+
\max{j\in\Lambda^C}~\left|\vw_j^T\ve\right|.
\label{eq:ZminCondInter2}
\end{equation}

\end{IEEEproof}

Note that so far we have made no specific assumptions on the signal generative model or the noise. The only assumption is on the inner products between the signal $\vx$ and rows in $\mOm$ that are not indexed in the true co-support. An immediate observation arising from the above lemma appears in the following corollary. Using the Cauchy-Schwarz inequality and the fact that all rows in $\mOm$ are normalized, we get that $\left|\vw_j^T\ve\right|\leq \|\ve\|_2$. Thus, 

\begin{corollary}
Let $\vy=\vx+\ve$, where $\vx$ is a co-sparse analysis signal with co-support $\Lambda$ on $\mOm$ and $\|\ve\|_2\leq\epsilon$. If $\vx$ and $\mOm$ satisfy $\zmin\geq 2\epsilon$, then the thresholding algorithm succeeds in recovering the true co-support $\Lambda$ of $\vx$ from $\vy$.
\label{corollary:THRAdversarialCond}
\end{corollary}

Note that we have referred to the noise as deterministic and bounded. This results in a very pessimistic success condition, as should be expected for a worst-case performance analysis like the one practiced here, in which an estimator must perform well even when the noise maximally damages the measurements (the noise in this case is thus called adversarial). This should remind the reader of the theoretical guarantees derived for synthesis-based pursuit algorithms under adversarial noise \cite{Review09,MyBook,Donoho01,Donoho03,Tropp04}.

To improve the theoretical guarantees, we turn to a setup where the noise is assumed to be random. Specifically, we assume white and zero-mean Gaussian noise with variance $\sigma^2$, and derive a lower bound on the probability of success under a sufficient condition on $\zmin$.

\begin{theorem}
Let $\vy=\vx+\ve$ and $\ve\sim N\left(0,\sigma^2\mI\right)$. If $\vx$ is a co-sparse analysis signal with co-support $\Lambda$ on $\mOm$, co-sparsity $\ell$, and co-rank $d-r$, and $\mOm$ and $\vx$ satisfy $\zmin\geq \beta\sigma$, then the thresholding algorithm succeeds in recovering the true co-support $\Lambda$ of $\vx$ from $\vy$ with probability at least $\left(\textnormal{Max}~\left\{0,1-\sqrt{\frac{8}{\pi\beta^2}}\exp\left\{-\frac{\beta^2}{8}\right\}\right\}\right)^{p-\ell+d-r}$.
\label{thm:THRGaussianCond}
\end{theorem}

\begin{figure}
\centering
\includegraphics[scale=0.4]{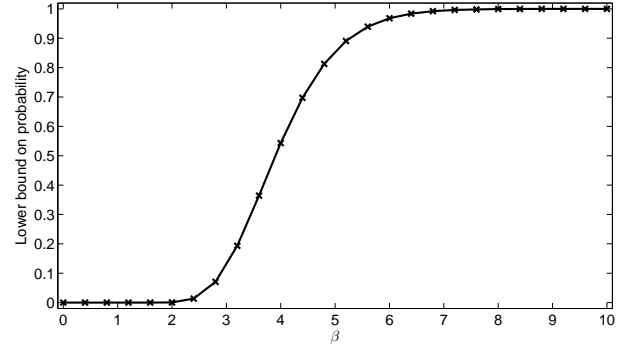}
\caption[]{The dependence on $\beta$ of the lower bound on the conditional probability of success given that $\zmin\geq \beta\sigma$ (see Theorem \ref{thm:THRGaussianCond}) for a setup with $d=9$, $p=18$, $r=2$ and $\ell=14$.}
\label{fig:DependenceOnBetaTheorem1}
\end{figure}

Before turning to prove this result, a short discussion is in order. This theorem provides a lower bound on the conditional probability of success given that $\zmin\geq\beta\sigma$. The derived expression has an exponential form with a base in the range $[0,1]$ depending on $\beta$ and a power $p-\ell+d-r$. In the rest of the paper we will denote the base of this exponential form by
\begin{equation}
g(\beta)=\textnormal{Max}~\left\{0,1-\sqrt{\frac{8}{\pi\beta^2}}\exp\left\{-\frac{\beta^2}{8}\right\}\right\}.
\label{eq:g}
\end{equation}
\noindent The observant reader might ask at this stage: Why is the performance guarantee of Theorem \ref{thm:THRGaussianCond} better than the result of Corollary \ref{corollary:THRAdversarialCond}? To answer this question we explore the dependence of this performance guarantee on $\beta$. The bound on this probability increases exponentially from zero to one as $\beta$ grows, but at the same time the condition on $\zmin$ becomes stricter. This bound is shown in Fig. \ref{fig:DependenceOnBetaTheorem1} for a setup with $d=9$, $p=18$, $r=2$ and $\ell=14$. First, we can see that the exact co-support is recovered with overwhelming probability (i.e. near one) for $\zmin\geq6\sigma$. This aligns with the guarantee of Corollary \ref{corollary:THRAdversarialCond} requiring $\zmin\geq2\epsilon$, where $\epsilon$ is of order $\sqrt{d}\sigma=3\sigma$. More importantly, Theorem \ref{thm:THRGaussianCond} provides probabilistic success guarantees for weaker conditions on $\zmin$, for which Corollary \ref{corollary:THRAdversarialCond} cannot make any guarantee.

Next, we explore the dependence of the obtained lower bound on the number of atoms $p$ and the co-sparsity $\ell$ and the co-rank $d-r$. Clearly, the probability of success of the thresholding algorithm improves (grows) when $p-\ell+d-r$ gets smaller. Such is the case, for example, when the dictionary size $(p,d)$ is kept fixed, the co-rank $d-r$ is chosen as well, and the level of dependencies, as depicted in $\ell$, grows. This manifests the surprising fact that strong linear-dependencies within $\mOm$ lead to better performance. Adopting a different point of view, when $p$ (the dictionary's redundancy) grows, the level of performance may remain the same as long as $\ell$ grows with it such that their difference remains unchanged.    

\begin{IEEEproof}
Let us first define the event
\begin{equation}
B=\left\{\ve~|~\max{j\in\widetilde{\Lambda}\cup\Lambda^C}~\left|\vw_j^T\ve\right|<\tau\right\}.
\label{eq:event}
\end{equation}
\noindent A similar event was defined in \cite{Ben-Haim10} when developing success guarantees for the synthesis-based thresholding and OMP algorithms. We start by deriving a lower bound on the probability of this event:
\begin{align}
\Pr\{B\}&\geq\prod\limits_{j\in\widetilde{\Lambda}\cup\Lambda^C}~\Pr\left\{\left|\vw_j^T\ve\right|<\tau\right\}=\left[1-2Q\left(\frac{\tau}{\sigma}\right)\right]^{p-\ell+d-r} \nonumber\\
&\geq\left[1-\sqrt{\frac{2\sigma^2}{\pi\tau^2}}\exp\left\{-\frac{\tau^2}{2\sigma^2}\right\}\right]^{p-\ell+d-r},
\label{eq:LowerBoundPrB}
\end{align}
\noindent where $Q(\cdot)$ is the Gaussian distribution tail,
\begin{equation}
Q(t)=\frac{1}{\sqrt{2\pi}}\int\limits_{t}^{\infty}\,\exp\left\{-\frac{z^2}{2}\right\}\mathrm{d}z.
\label{eq:GaussianTail}
\end{equation}
\noindent The first inequality holds due to \^{S}id\'{a}k's lemma \cite{Sidak67} for a set of jointly Gaussian random variables. The next equality holds due to the fact that $\widetilde{\Lambda}$ and $\Lambda^C$ are disjoint sets of sizes $d-r$ and $p-\ell$ respectively. In the last inequality we use a well-known upper bound on the Gaussian distribution tail,
\begin{equation}
Q(t)\leq\frac{1}{t\sqrt{2\pi}}\exp\left\{-\frac{t^2}{2}\right\}.
\label{eq:GaussianUpperBound}
\end{equation}

We set $\tau=\frac{1}{2}\beta\sigma$, and thus the event $B$ corresponds to all the noise vectors $\ve$ satisfying $2\text{Max}_{j\in\widetilde{\Lambda}\cup\Lambda^C}~\left|\vw_j^T\ve\right|<\beta\sigma$. Therefore, if $\zmin>\beta\sigma$ as this theorem states, then necessarily $\zmin$ also satisfies the condition of Lemma \ref{lemma:THRGeneralCond}, namely $\zmin>\beta\sigma>2\text{Max}_{j\in\widetilde{\Lambda}\cup\Lambda^C}~\left|\vw_j^T\ve\right|$, which guarantees the success of the analysis thresholding algorithm. The probability for this to happen is bounded from below by the expression we have derived in Eq. (\ref{eq:LowerBoundPrB}), as claimed\footnote{For values of $\beta$ that lead to a negative argument in this expression we replace Eq. (\ref{eq:LowerBoundPrB}) by a trivial zero lower bound on the probability.}.

\end{IEEEproof}

Next, we would like to eliminate the dependence on $\zmin$ and derive a theoretical guarantee in terms of the analysis subspace dimension $r$, the co-sparsity $\ell$ and possibly some internal properties of the dictionary $\mOm$. This will help to reveal what makes an analysis dictionary more suitable for co-sparse estimation. To initiate such an analysis, we make an additional assumption on the signal generative model. Given a dictionary $\mOm$, a co-support $\Lambda$ and a random Gaussian vector $\vu\sim N\left(0,\sigma_u^2\mI\right)$, $\vx$ is generated by projecting $\vu$ onto the subspace orthogonal to $\mOm_{\Lambda}$, as described in Section \ref{subsec:THRExperiments} (see (\ref{eq:analysis-projection})). We further assume that $\vu$ and $\ve$ are statistically independent. Using this generative model for $\vx$, we shall derive a theoretical guarantee for success of the thresholding algorithm, based on a new property of $\mOm$ we shall refer to as ROPP: 

\begin{definition} Given an analysis dictionary $\mOm$, the Restricted Orthogonal Projection Property (ROPP) of this dictionary with a constant $\alpha_r$ is defined as
\begin{equation}
\alpha_r=\min{\Lambda,j|\rank\left(\mOm_\Lambda\right)=d-r, j\in\Lambda^C}~\|(\mI-\mOm_\Lambda^\dagger\mOm_\Lambda)\vw_j\|_2.
\label{eq:ROPP}
\end{equation}
\vspace{-20pt}
\label{def:ROPP}
\end{definition}

\noindent More on the meaning of this constant is brought in Section \ref{subsec:Discussion}. Armed with this definition, we now turn to improve Theorem \ref{thm:THRGaussianCond}, by removing the dependency on $\zmin$.

\begin{theorem}
Let $\vy=\vx+\ve$, where $\vu\sim N\left(0,\sigma_u^2\mI\right)$, $\vx$ is a co-sparse analysis signal with co-support $\Lambda$ on $\mOm$, obtained by $\vx=(\mI-\mOm_\Lambda^\dagger\mOm_\Lambda)\vu$, and $\ve\sim N\left(0,\sigma^2\mI\right)$ is the additive noise statistically independent of $\vu$. If $\mOm$ satisfies the ROPP with a constant $\alpha_r$ and $\vx$ has co-rank $d-r$ and co-sparsity $\ell$ on $\mOm$, then the thresholding algorithm succeeds in recovering the true co-support $\Lambda$ of $\vx$ from $\vy$ with probability at least $\left[g(\beta)\right]^{p-\ell+d-r}\left(2Q\left(\frac{\beta\sigma}{\alpha_r\sigma_u}\right)\right)^{p-\ell}$ for any constant $\beta>0$.
\label{thm:AlphaEllCond}
\end{theorem}
\noindent Note that the function $g(\cdot)$ appearing in this theorem is defined in Eq. (\ref{eq:g}) and $Q(\cdot)$ is the Gaussian distribution tail (see Eq. (\ref{eq:GaussianTail})).

\begin{figure}
\centering
\vspace{-10pt}
\includegraphics[scale=0.4]{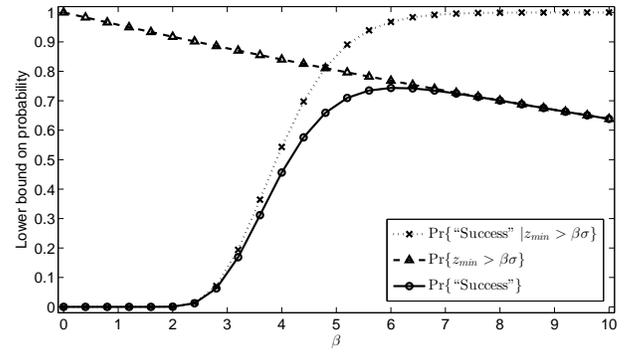}
\vspace{-10pt}
\caption[]{The dependence on $\beta$ of the lower bound on the probability of success of Theorem \ref{thm:AlphaEllCond} for a setup with $d=9$, $p=18$, $r=2$, $\ell=14$, $\sigma/\sigma_u=0.01$ and $\alpha_r=0.75$. For this setup the optimal value of $\beta$ is $6$, which results in a lower bound of $0.744$ on the probability of success. For each value of $\beta$ we also show the lower bounds on the conditional probability of success of Theorem \ref{thm:THRGaussianCond} and on the probability that the condition $\zmin\geq\beta\sigma$ holds (see Eq. (\ref{eq:PrZminCond})). The final bound of Theorem \ref{thm:AlphaEllCond} is a product of these two bounds.}
\label{fig:DependenceOnBetaTheorem2}
\end{figure}

Just as we did for the conditional probability of success of Theorem \ref{thm:THRGaussianCond}, we start by exploring the dependence of the resulting bound with respect to $\beta$. This is shown in Fig. \ref{fig:DependenceOnBetaTheorem2} for a setup with $d=9$, $p=18$, $r=2$, $\ell=14$ (same as before -- see Fig. \ref{fig:DependenceOnBetaTheorem1}), $\alpha_r=0.75$ and $\sigma/\sigma_u=0.01$. We can see that the choice of $\beta$ is crucial for the strictness of the resulting lower bound on the probability of success. For the setup considered here the optimal value of $\beta$ is $6$, which results in a lower bound of $0.744$. The lower bound appearing in this theorem is a product of two exponential terms. The first is the bound on the conditional probability that appeared in Theorem \ref{thm:THRGaussianCond} and the second terms is a bound on the probability that the condition $\zmin\geq\beta\sigma$ holds (this bound will be derived in the proof that follows). The first terms grows with $\beta$, while the second decreases, thus explaining the peak between $0$ and infinity.  

Next, we explore the dependence of the obtained lower bound on the number of atoms $p$ and the co-sparsity $\ell$, fixing the noise ratio $\sigma/\sigma_u$, the signal dimension $d$ and the analysis subspace dimension $r$, and assuming that the dictionary satisfies the ROPP with a constant $\alpha_r$. Since both the bases of the exponential terms are in the range $[0,1]$, we can see that the probability of success of the thresholding algorithm improves when the difference $p-\ell$ becomes smaller. This means that the same observations made before on $p$ and $\ell$ for the conditional probability also hold here: For a given dictionary of size $(p,d)$ performance improves as $\ell$ grows, and when the redundancy of the dictionary is increased the performance remain the same as long as the difference $p-\ell$ remains unchanged. Finally, we observe that since $Q(\cdot)$ is monotonic decreasing, the performance improves as the noise ratio $\sigma/\sigma_u$ decreases or the ROPP constant $\alpha_r$ grows. 

\begin{IEEEproof}
We begin by observing that a signal $\vx$ generated as an orthogonal projection of a Gaussian i.i.d. vector $\vu$ is also Gaussian, $\vx\sim N\left(0,\sigma_u^2(\mI-\mOm_\Lambda^\dagger\mOm_\Lambda)\right)$ and so is any inner product with $\vx$, $\vw_j^T\vx\sim N\left(0,\|(\mI-\mOm_\Lambda^\dagger\mOm_\Lambda)\vw_j\|_2^2\sigma_u^2\right)$. Using this observation, we now derive a lower bound on the probability that the condition for success of Theorem \ref{thm:THRGaussianCond} holds:
\begin{align}
&\Pr\left\{\zmin>\beta\sigma\right\}=\Pr\left\{\min{j\in\Lambda^C}~\left|\vw_j^T\vx\right|>\beta\sigma\right\}\nonumber\\
&\geq\prod\limits_{j\in\Lambda^C}~\Pr\left\{\left|\vw_j^T\vx\right|>\beta\sigma\right\} \label{eq:PrZminCond}\\
&=\prod\limits_{j\in\Lambda^C}~2Q\left(\frac{\beta\sigma}{\|(\mI-\mOm_\Lambda^\dagger\mOm_\Lambda)\vw_j\|_2\sigma_u}\right)
\geq\left[2Q\left(\frac{\beta\sigma}{\alpha_r\sigma_u}\right)\right]^{p-\ell}\nonumber.
\end{align}
\noindent The first inequality relies on \^{S}id\'{a}k's lemma, as before\footnote{In fact, we are not explicitly using \^{S}id\'{a}k's lemma, but a related inequality resulting from this lemma. Let $\{v_j\}_{j=1}^M$ be a set of jointly Gaussian random vectors. Then according to \^{S}id\'{a}k's lemma, $\Pr\left\{\text{Max}_{1\leq j\leq M}~|v_j|<\tau\right\}\geq\prod\limits_{j=1}^M~\Pr\left\{|v_j|<\tau\right\}$. Thus, turning to our expression, we observe that $\Pr\left\{\text{Min}_{1\leq j\leq M}~|v_j|>\tau\right\}=\Pr\left\{-\text{Max}_{1\leq j\leq M}~(-|v_j|)>\tau\right\}=\Pr\left\{\text{Max}_{1\leq j\leq M}~(-|v_j|)<-\tau\right\} \geq\prod\limits_{j=1}^M~\Pr\left\{-|v_j|<-\tau \right\}=\prod\limits_{j=1}^M~\Pr\left\{|v_j|>\tau \right\}$, leading to the relation we used.}. In the next equality we use the fact that $\vw_j^T\vx$ is Gaussian with the variance mentioned above. The last inequality holds from the definition of the ROPP in (\ref{eq:ROPP}) and since $Q(\cdot)$ is monotonic decreasing. The power $p-\ell$ comes from the cardinality of the set $\Lambda^C$. 

Combining Theorem \ref{thm:THRGaussianCond} and Eq. (\ref{eq:ZminCond}) we get that the final lower bound on the probability of success for the thresholding algorithm is a direct multiplication of the two probability expressions, leading to the claimed lower-bound probability posed in terms of the ROPP constant $\alpha_r$ and the co-sparsity $\ell$.
\end{IEEEproof}

\subsection{Discussion on the Properties of the Analysis Dictionary}

\label{subsec:Discussion}

We begin this subsection by taking a closer look at the ROPP. This is an internal property of the analysis dictionary, indicating for a set of $d-r+1$ linearly independent rows from the dictionary how much each row is spread away from the subspace spanned by the rest. At the special case of a unitary dictionary $\mOm$ we have $\alpha_r=1$ for all values of $r$ since each row is orthogonal to the subspace spanned by every possible set of rows not including it. How does the ROPP compares to other dictionary properties? Starting with the RIP \cite{Candes05,Candes06},
\begin{equation}
\left(1-\delta_k\right)\|\vv\|_2^2\leq\|\mD\vv\|_2^2\leq \left(1+\delta_k\right)\|\vv\|_2^2,
\label{eq:RIP}
\end{equation}
\noindent which holds for all $k$-sparse vectors $\vv\in\RR^n$, the ROPP also bounds an $\ell_2$ norm related to the dictionary. However, the ROPP looks at projection matrices constructed from the dictionary instead of the dictionary itself as in the RIP, and applies these matrices on dictionary atoms not used for the matrix construction instead of looking at all possible signals with a certain sparsity as in the RIP. This should remind the reader of the ERC \cite{Tropp04}, which has a similar flavor. Turning to the ERC \cite{Tropp04}, for a better comparison let us replace the ROPP by the sufficient condition
\begin{equation}
\max{j\in\Lambda^C}{\|\mOm_\Lambda^\dagger\mOm_\Lambda\vw_j\|_2}\leq 1-\alpha_r
\label{eq:SufficientROPP}
\end{equation}
\noindent for the same co-supports $\Lambda$ as in (\ref{eq:ROPP}). To see that this is indeed a sufficient condition, we assume that (\ref{eq:SufficientROPP}) holds and show that 
\begin{equation}
\|\left(\mI-\mOm_\Lambda^\dagger\mOm_\Lambda\right)\vw_j\|_2\geq\|\vw_j\|_2-\|\mOm_\Lambda^\dagger\mOm_\Lambda\vw_j\|_2\geq\alpha_r,
\label{eq:SufficientROPPJustified}
\end{equation}
\noindent where in the first inequality we used the well-known relation, $\|\vv_1-\vv_2\|_2\geq\left|\|\vv_1\|_2-\|\vv_2\|_2\right|$, which holds for any pair of vectors $\vv_1,\vv_2$, and in the second inequality we used the fact that $\|\vw_j\|_2=1$ and the assumption of (\ref{eq:SufficientROPP}). The condition appearing in (\ref{eq:SufficientROPP}) has a similar structure to the ERC,
\begin{equation}
\max{j\notin s}{\|\mD^\dagger_s\vd_j\|_1}\leq 1.
\label{eq:ERC}
\end{equation}
\noindent However, there are two inherent differences: The pseudoinverse of the submatrix $\mD_s$ is replaced by a projection matrix onto the null space of $\mOm_\Lambda$ and the $\ell_1$ norm is replaced by $\ell_2$. Consequently, an upper bound of $1$ is a trivial one and it is replaced by the stricter bound $1-\alpha_r$ for some constant $\alpha_r$.

\begin{figure}
\centering
\includegraphics[scale=0.4]{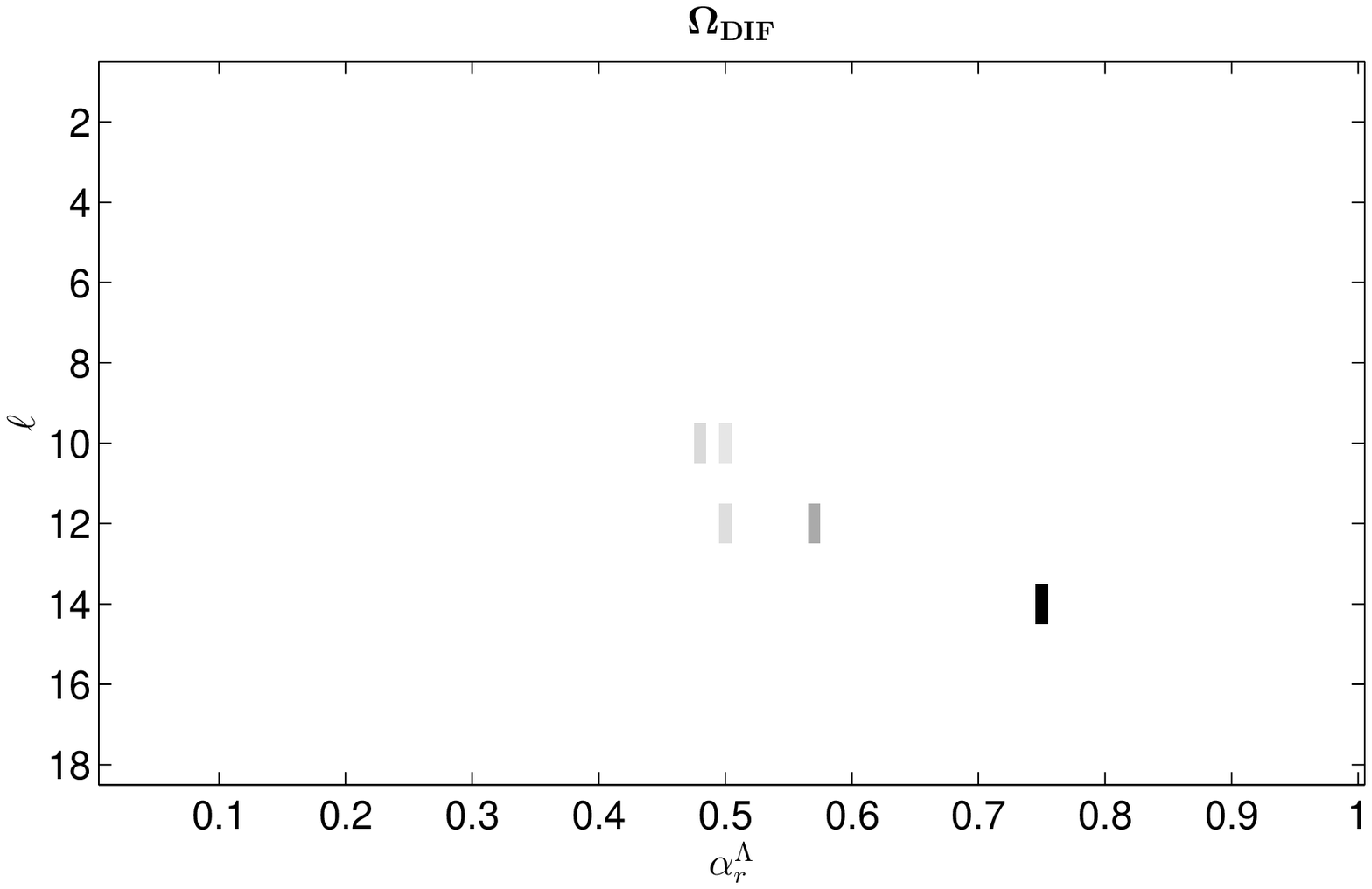}\\
\vspace{10pt}
\includegraphics[scale=0.4]{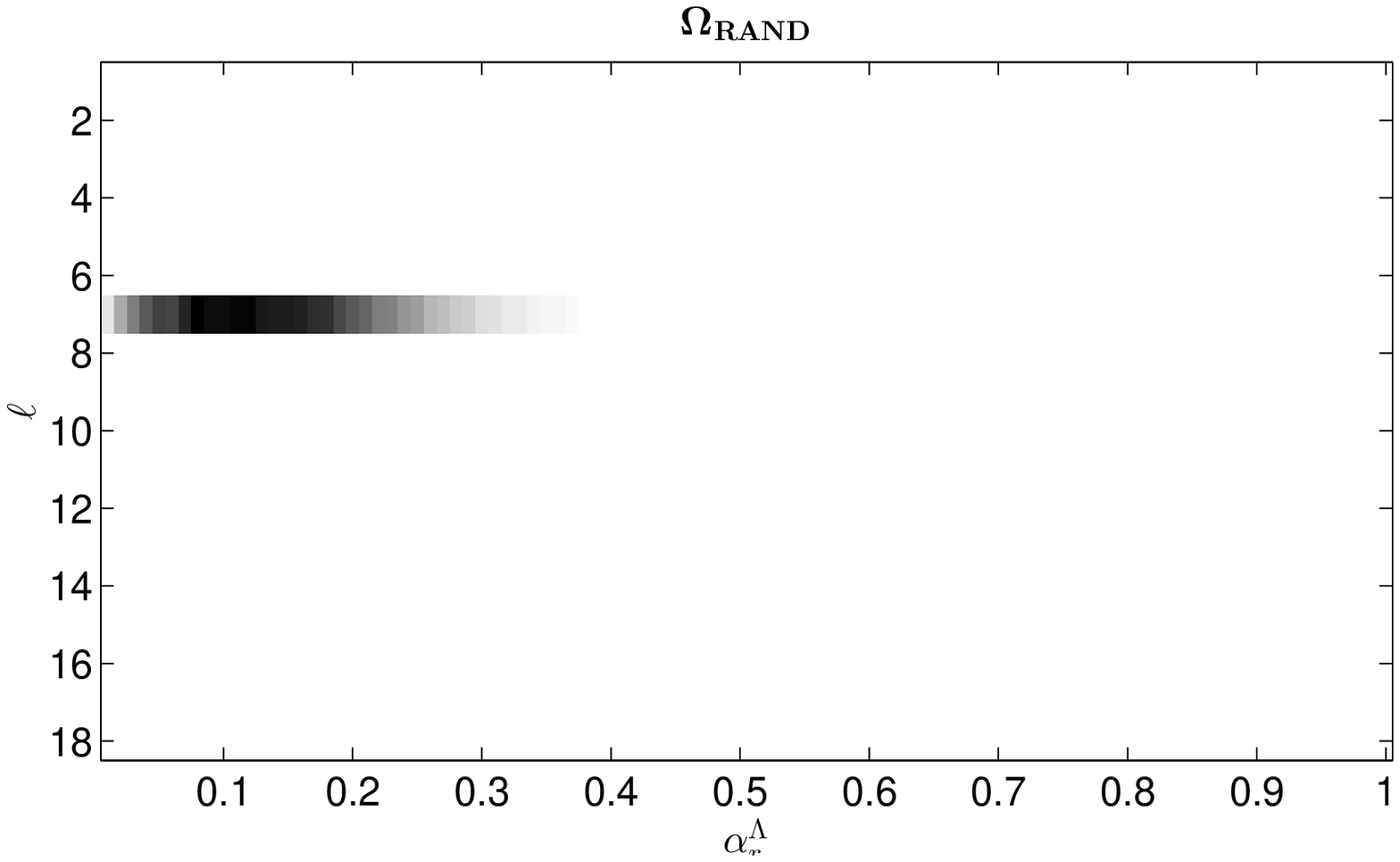}\\
\vspace{10pt}
\includegraphics[scale=0.4]{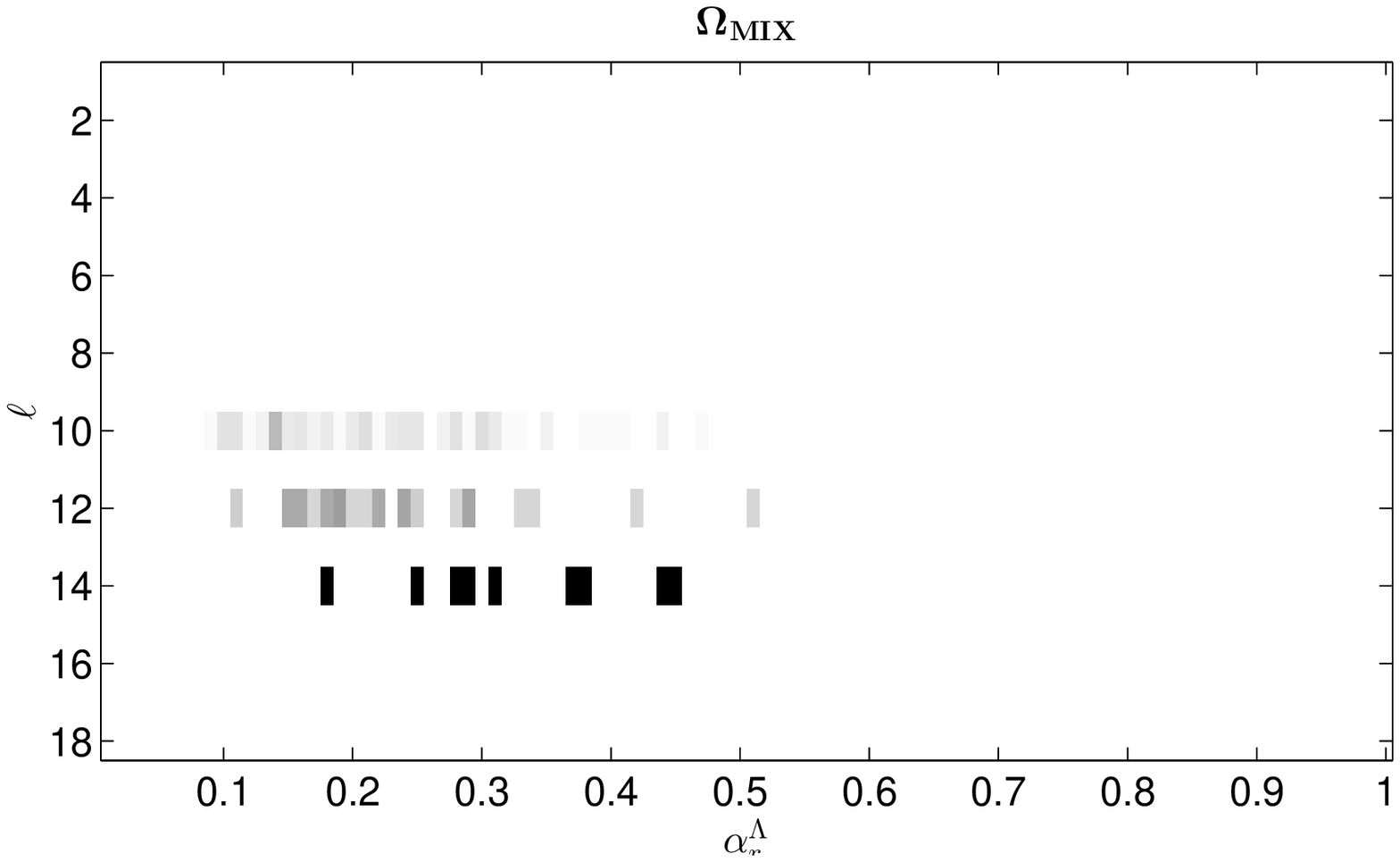}
\caption[]{The joint distribution of $\ell$ and $\alpha_r^\Lambda$ for each type of the analysis dictionaries of size $18\times 9$ that were shown in Fig. \ref{fig:OmegaTypes} and for $r=2$. Each of these distributions is obtained by an exhaustive computation over all possible subsets of rows from the analysis dictionary with co-rank $7$, and is displayed in the form of a matrix $\mP^{(2)}$, whose entries where defined in Eq. (\ref{eq:JointDist}). A darker bin corresponds to a higher value in the joint distribution.}
\label{fig:JointDistEllAlpha}
\end{figure}

Next, we turn to the theoretical guarantee of Theorem \ref{thm:AlphaEllCond} and observe that it gives rise to two dictionary properties, which serve as two distinct forces dictating the ability to recover the co-supports of analysis signals over the given dictionary. The first property, emanating from the signature or the co-sparsity of $\mOm$, determines which sets of rows and how many of them are linearly dependent. However, this measure by itself does not provide us with any quantitative relation between these sets and the rows that are linearly independent on them. The second property focuses exactly on these missing relations, telling us how much a row is spread away from the others, provided that it is linearly independent on them.

Are these two dictionary properties somehow related to each other? To provide an answer to this question we explore the joint distribution of the two. For this purpose, we replace $\alpha_r$ by $\alpha_r^\Lambda$ which has a similar definition, apart from a delicate modification: It should satisfy (\ref{eq:ROPP}) for a {\em single} co-support $\Lambda$ corresponding to a co-rank $d-r$, rather than for all possible co-supports leading to this co-rank, as in the definition of $\alpha_r$ (see Definition \ref{def:ROPP}). This means that $\alpha_r$ can be obtained by taking the minimal value of $\alpha_r^\Lambda$ over all of these co-supports. Since $\alpha_r^\Lambda$ is a continuous measure in the range $[0,1]$, and since we are about to create histograms of possible values, we perform a uniform quantization of $\alpha_r^\Lambda$ to $T=100$ discrete levels. The joint distribution of $\ell$ and $\alpha_r^\Lambda$ is represented by a $p$-by-$T$ matrix with entries
\begin{equation}
P^{(r)}_{km}=\Pr\left\{\ell=k,\frac{m-1}{T}\leq\alpha_r^\Lambda<\frac{m}{T}\right\},
\label{eq:JointDist}
\end{equation}
\noindent Obtaining the entries of the matrix $\mP^{(r)}$ requires an exhaustive computation over all possible co-supports with co-rank $d-r$. The joint distributions for the three dictionaries (shown in Fig. \ref{fig:OmegaTypes}) and a co-rank of $7$ (i.e. $r=2$) are depicted in Fig. \ref{fig:JointDistEllAlpha}. We can see that increasing the co-sparsity level typically spreads $\alpha_r^\Lambda$ towards higher values. This makes sense since the minimization appearing in (\ref{eq:ROPP}) is performed over smaller index sets.

\subsection{Results of the Analysis Thresholding Revisited}

\label{subsec:THRPerformanceRevisited}

We revisit the results shown in Section \ref{subsec:THRExperiments} and try to explain them in light of the theoretical guarantees derived in Section \ref{subsec:THRGuarantees}. Note that the setup considered in Theorem \ref{thm:AlphaEllCond} (projection of a white Gaussian vector $\vu$, additive white Gaussian noise) matches completely the one used for the experiments of Section \ref{subsec:THRExperiments}. This will allow us to make the desired connections between the empirical results and the theoretical guarantee. An immediate observation arising from Theorem \ref{thm:AlphaEllCond} is that the higher the co-sparsity level $\ell$ of $\vx$ with respect to $\mOm$, the better the thresholding algorithm is expected to perform in recovering the true co-support. This implies that linear dependencies within $\mOm$ are highly desired. This stands as a complete contradiction to the intuition gained for the synthesis-based sparsity model, where such dependencies between the atoms lead to a collapse of pursuit algorithms. We also observe that the results of the analysis thresholding algorithm improve as $\alpha_r$ grows. This is closer in spirit to the ERC/RIP rationale, where independencies are encouraged.

\begin{figure}
\centering
\includegraphics[scale=0.5]{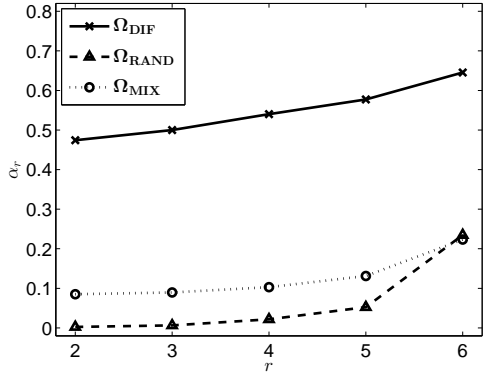}
\caption[]{The values of the ROPP constant for each type of the analysis dictionaries of size $18\times 9$ that were shown in Fig. \ref{fig:OmegaTypes} and for varying analysis subspace dimensions $r$. Each of these values is obtained by an exhaustive minimization over all possible subsets of rows from the analysis dictionary with co-rank $9-r$.}
\label{fig:OmegaROPP}
\end{figure}

Returning to the empirical results of Section \ref{subsec:THRExperiments}, we have already seen in Fig. \ref{fig:OmegaCoSparsity} that $\mOm_{DIF}$ and $\mOm_{MIX}$ have the same co-sparsity distribution, where the co-sparsity can be much higher than the co-rank $d-r$. This can explain, at least in part, their superior performance over $\mOm_{RAND}$, which allows only a constant co-sparsity level $\ell=d-r$. We now turn to examine the value of the ROPP constant for each type of dictionary, with a hope to reveal an additional inherent difference between the dictionaries. These values are shown in Fig. \ref{fig:OmegaROPP} for the three dictionary types and for varying analysis subspace dimensions $r$. To obtain each of these values we performed an exhaustive minimization over all possible subsets $\Lambda$ of rows from $\mOm$ such that $\rank\left\{\mOm_\Lambda\right\}=d-r$. We can see that $\mOm_{DIF}$ corresponds to a much higher ROPP constant for all the examined co-ranks, when compared to $\mOm_{MIX}$ and $\mOm_{RAND}$. The two latter dictionaries have very low ROPP constants (below $0.14$ for $r\leq 5$). Specifically, at a subspace dimension of $r=2$ that was considered in the experiments of Section \ref{subsec:THRExperiments}, the ROPP constant is $5.6$ times higher for $\mOm_{DIF}$ compared to $\mOm_{MIX}$ and $202$(!) times higher compared to $\mOm_{RAND}$. We can conclude that the value of the ROPP constant explains the superior behavior of the thresholding algorithm with $\mOm_{DIF}$ when compared to $\mOm_{MIX}$, as observed in Fig. \ref{fig:EmpiricalResultsTHR}. This dictionary property also provides additional grounds for the inferior behavior with $\mOm_{RAND}$.

\begin{figure*}
\centering
\includegraphics[scale=0.5]{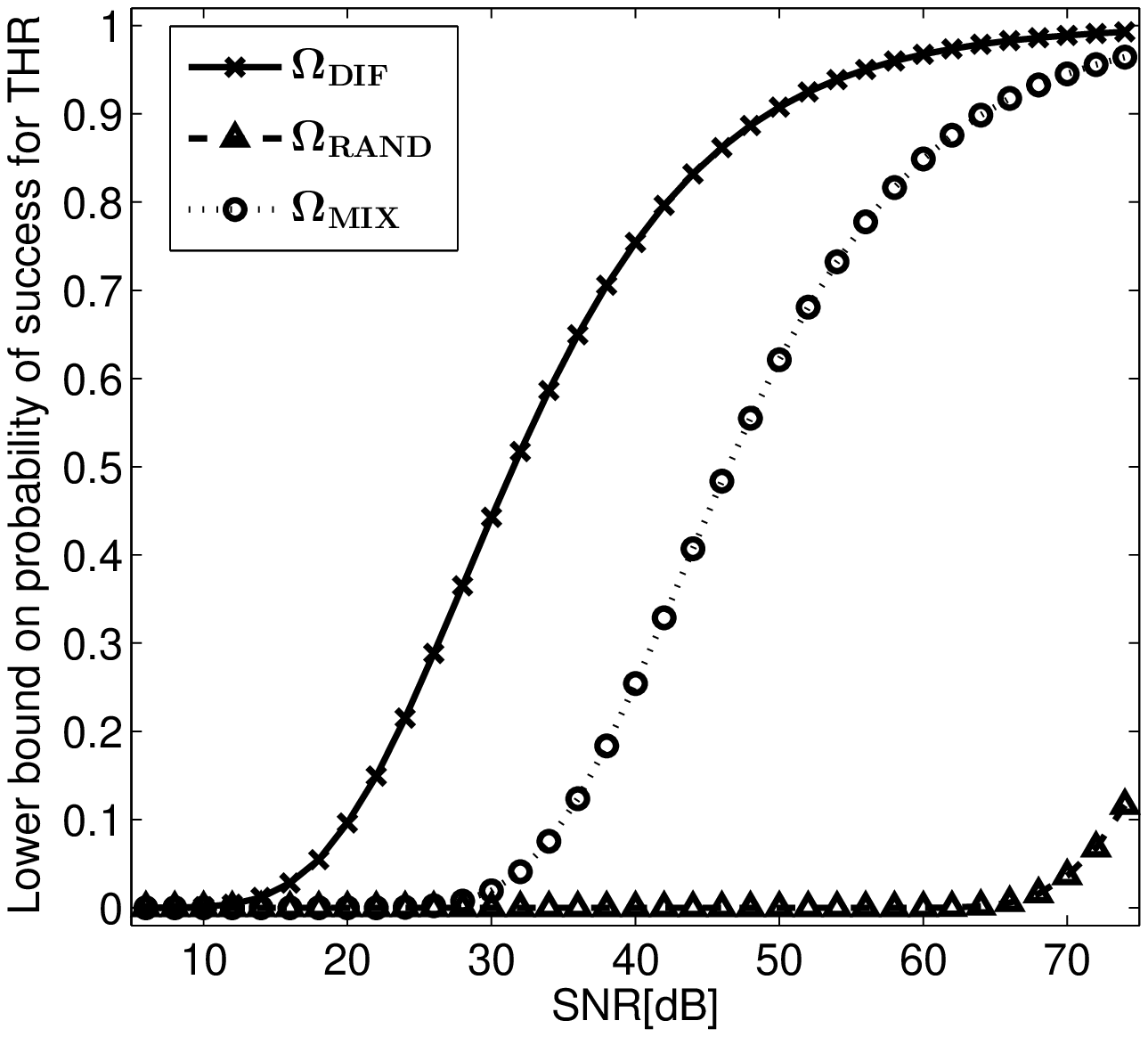}
\hspace{10pt}
\includegraphics[scale=0.5]{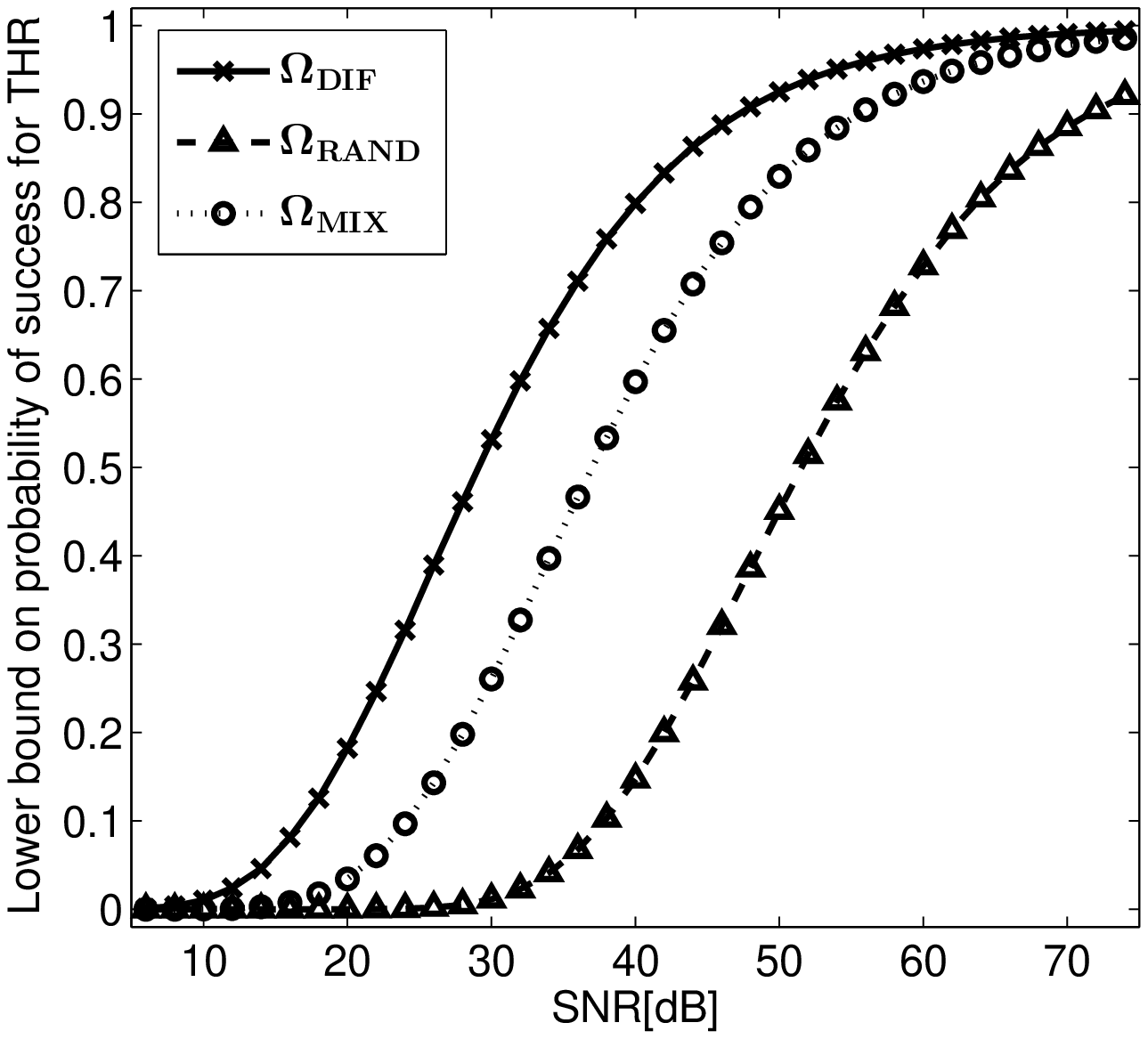}
\caption[]{Lower bounds on the probability of success for the thresholding algorithm on the three types of analysis dictionaries of size $18\times 9$ that were shown in Fig. \ref{fig:OmegaTypes} and for varying SNR levels. Top: For each ratio $\sigma/\sigma_u$ a lower bound is computed using Eq. (\ref{eq:LowerBoundPrSuccess}), where for each co-sparsity level $\ell$ we choose a value for $\beta$ such that the resulting bound will be as tight as possible. Bottom: For each ratio $\sigma/\sigma_u$ a lower bound is computed using Eq. (\ref{eq:TighterLowerBoundPrSuccess}), where an optimal value for $\beta$ is set for each pair $\ell,~\alpha_r^\Lambda$. As can be seen, the bounds appearing on the right are tighter than those shown on the left.}
\label{fig:TheoreticalPrSuccess}
\end{figure*}

Next, we turn to examine the theoretical success guarantee provided in Theorem \ref{thm:AlphaEllCond}. Fig. \ref{fig:TheoreticalPrSuccess} (top) displays this lower bound on the probability of success for the thresholding algorithm for each of the dictionaries and for varying SNR levels in the range $6dB$ to $74dB$ \footnote{See Eq. (\ref{eq:SNR computation}) for the definition of SNR and its dependence on $\sigma/\sigma_u$.}. To obtain each of the lower bounds that are shown in this figure, we find for each co-sparsity $\ell$ and each noise ratio $\sigma/\sigma_u$ a value of $\beta$ such that the lower bound for the probability of success provided in Theorem \ref{thm:AlphaEllCond} is as tight (i.e. high) as possible. An example of how to choose an optimal value of $\beta$ was depicted in Fig. \ref{fig:DependenceOnBetaTheorem2}. Finally, we perform a weighted average of these lower bounds, where the weights are simply the values of the co-sparsity distribution. This process can be described by the following equation:
\begin{align}
&\Pr\{\textnormal{``Success''}\}=\sum\limits_{k=1}^p~\Pr\{\ell=k\}\Pr\{\textnormal{``Success''}|\ell=k\} \nonumber \\
&\geq\sum\limits_{k=1}^p~\Pr\{\ell=k\}\left[g\left(\beta_k\right)\right]^{p-k+d-r}\left[2Q\left(\frac{\beta_k\sigma}{\alpha_r\sigma_u}\right)\right]^{p-k},
\label{eq:LowerBoundPrSuccess}
\end{align}
\noindent where the function $g(\cdot)$ is defined in Eq. (\ref{eq:g}) and $\beta_k$ is the value of $\beta$ that is set for co-sparsity $\ell=k$. These values are chosen such that the arguments inside the sum are maximized for each $k$ separately.

We can see that the resulting lower bounds can provide some insight into the actual performance. They are capable of predicting success with high probability at high SNR levels for $\mOm_{DIF}$ and $\mOm_{MIX}$. Another useful property of these bounds is that they clearly predict which dictionary the thresholding algorithm is expected to perform better with and which would probably lead to failure. Note that in our quest for theoretical guarantees we have lost much tightness with respect to the empirical results. This is typical for a theoretical analysis, but as we shall see in a moment, the tightness of the derived bounds can be considerably improved if we take into account the fact that $\alpha_r^\Lambda$ varies as a function of the co-support, and has a spread of values. Specifically, we can modify the process described in Eq. (\ref{eq:LowerBoundPrSuccess}) by replacing the distribution of $\ell$ and the fixed worst-case value of $\alpha_r$ with the joint distribution of $\ell$ and $\alpha_r^\Lambda$, as depicted in Fig. \ref{fig:JointDistEllAlpha}. For each such pair and for each noise ratio $\sigma/\sigma_u$ we set an optimal value of $\beta$ as described before, and use the values of the joint distribution as weights for the final average. This means that the process of (\ref{eq:LowerBoundPrSuccess}) is replaced by
\begin{align}
&\Pr\{\textnormal{``Success''}\} \nonumber\\
&=\sum\limits_{k=1}^p\sum\limits_{m=1}^T~P_{km}^{(r)}\Pr\left\{\textnormal{``Success''}|\ell=k,\frac{m-1}{T}\leq\alpha_r^\Lambda<\frac{m}{T}\right\} \nonumber \\
&\geq\sum\limits_{k=1}^p\sum\limits_{m=1}^T~P_{km}^{(r)}\left[g\left(\beta_{km}\right)\right]^{p-k+d-r}\left[2Q\left(\frac{\beta_{km}T\sigma}{(m-1)\sigma_u}\right)\right]^{p-k}.
\label{eq:TighterLowerBoundPrSuccess}
\end{align}
\noindent The resulting lower bounds are shown on the bottom of Fig. \ref{fig:TheoreticalPrSuccess} and as can be seen, they are much tighter than the previous ones appearing in this figure on the top.

\begin{figure*}
\centering
\includegraphics[scale=0.4]{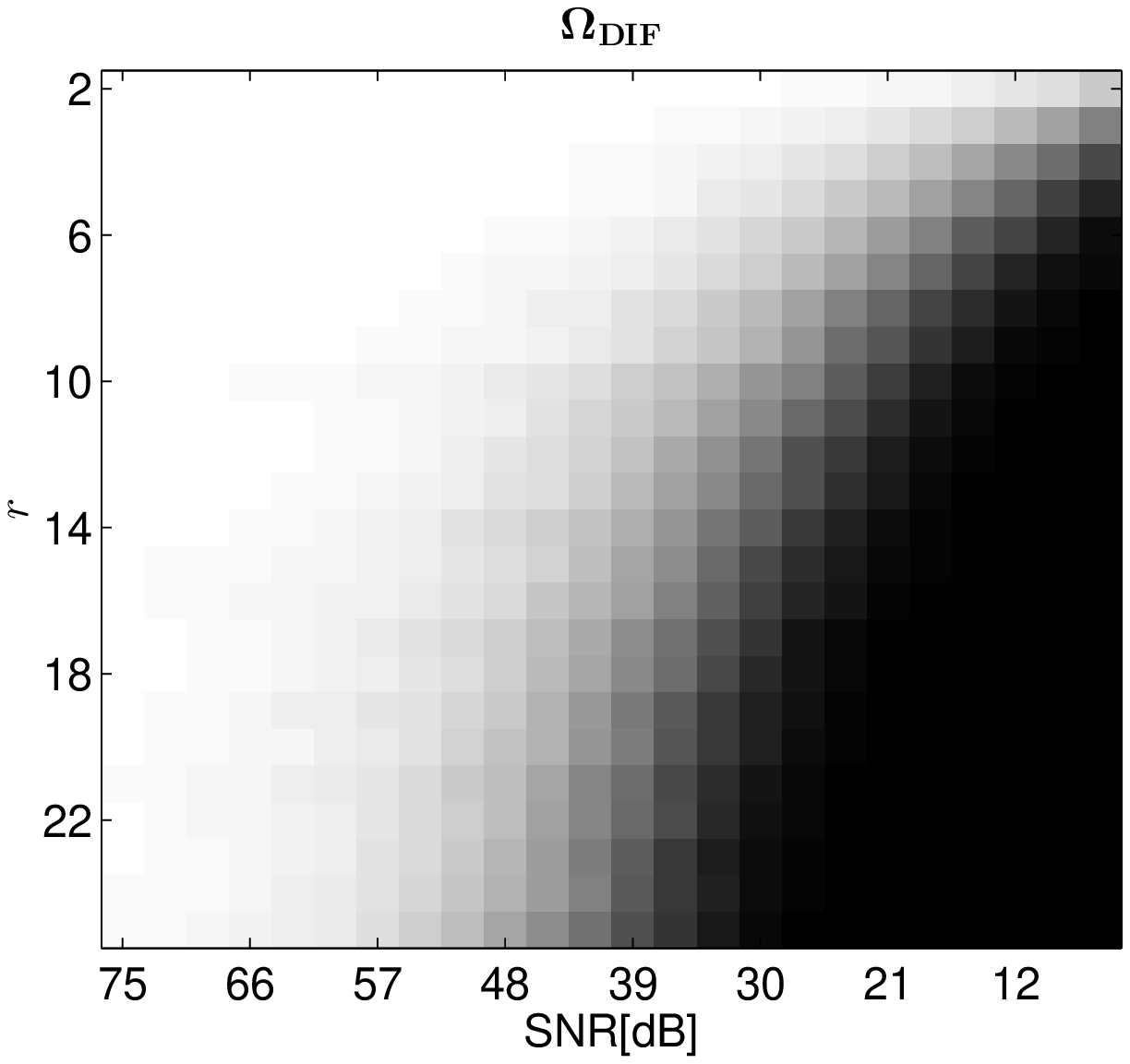} \hspace{20pt}
\includegraphics[scale=0.4]{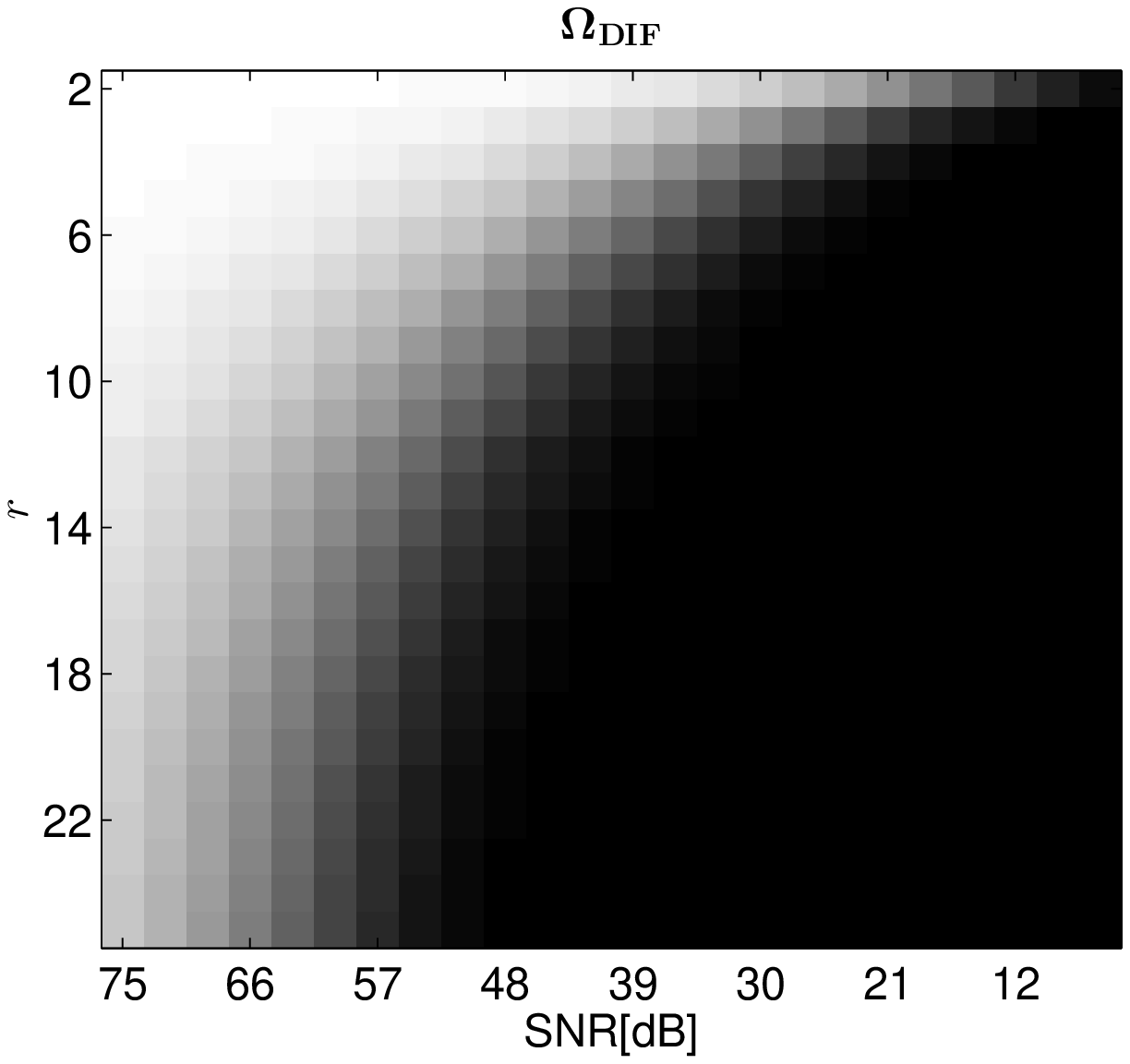}\\
\vspace{10pt}
\includegraphics[scale=0.4]{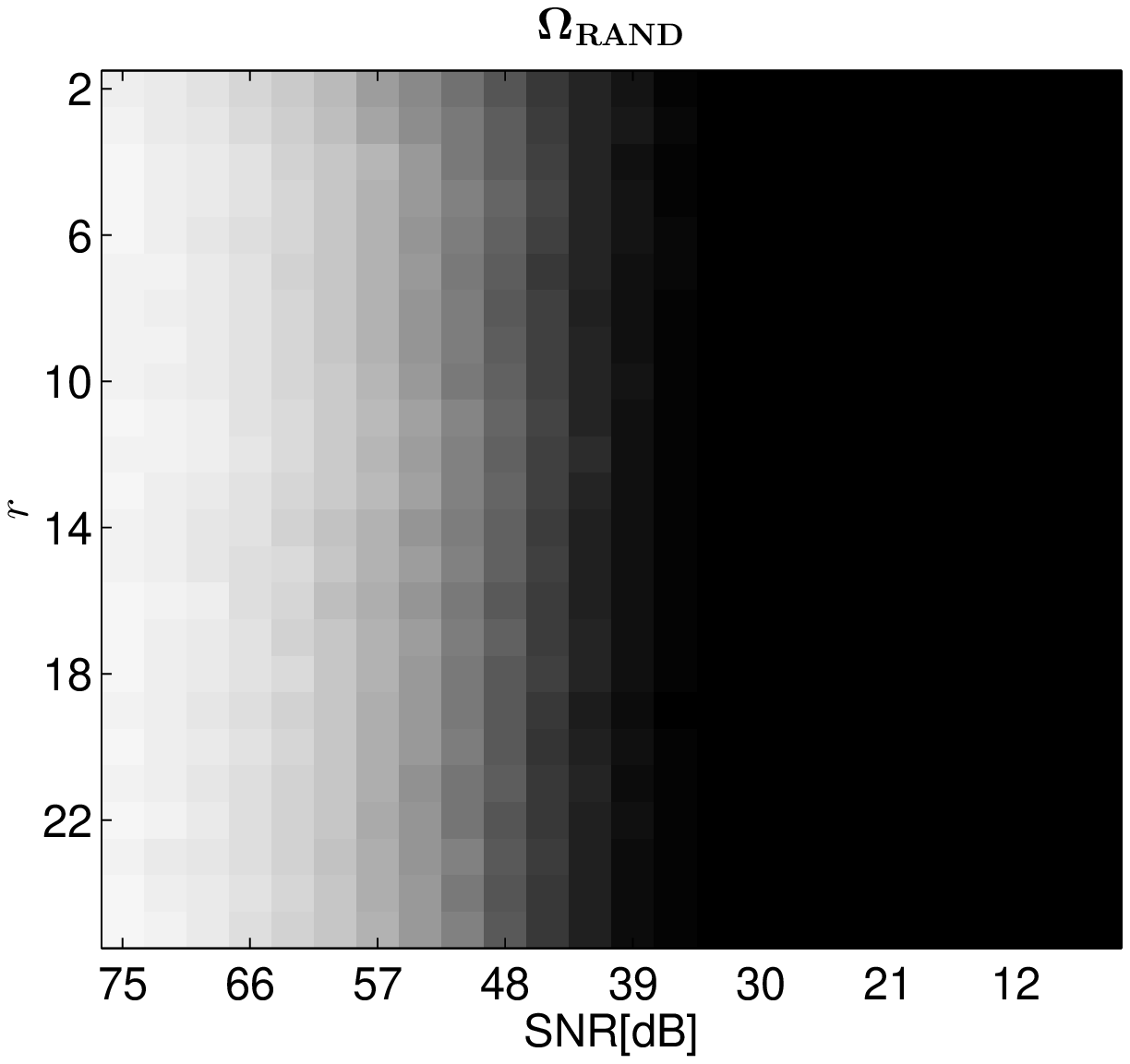} \hspace{20pt}
\includegraphics[scale=0.4]{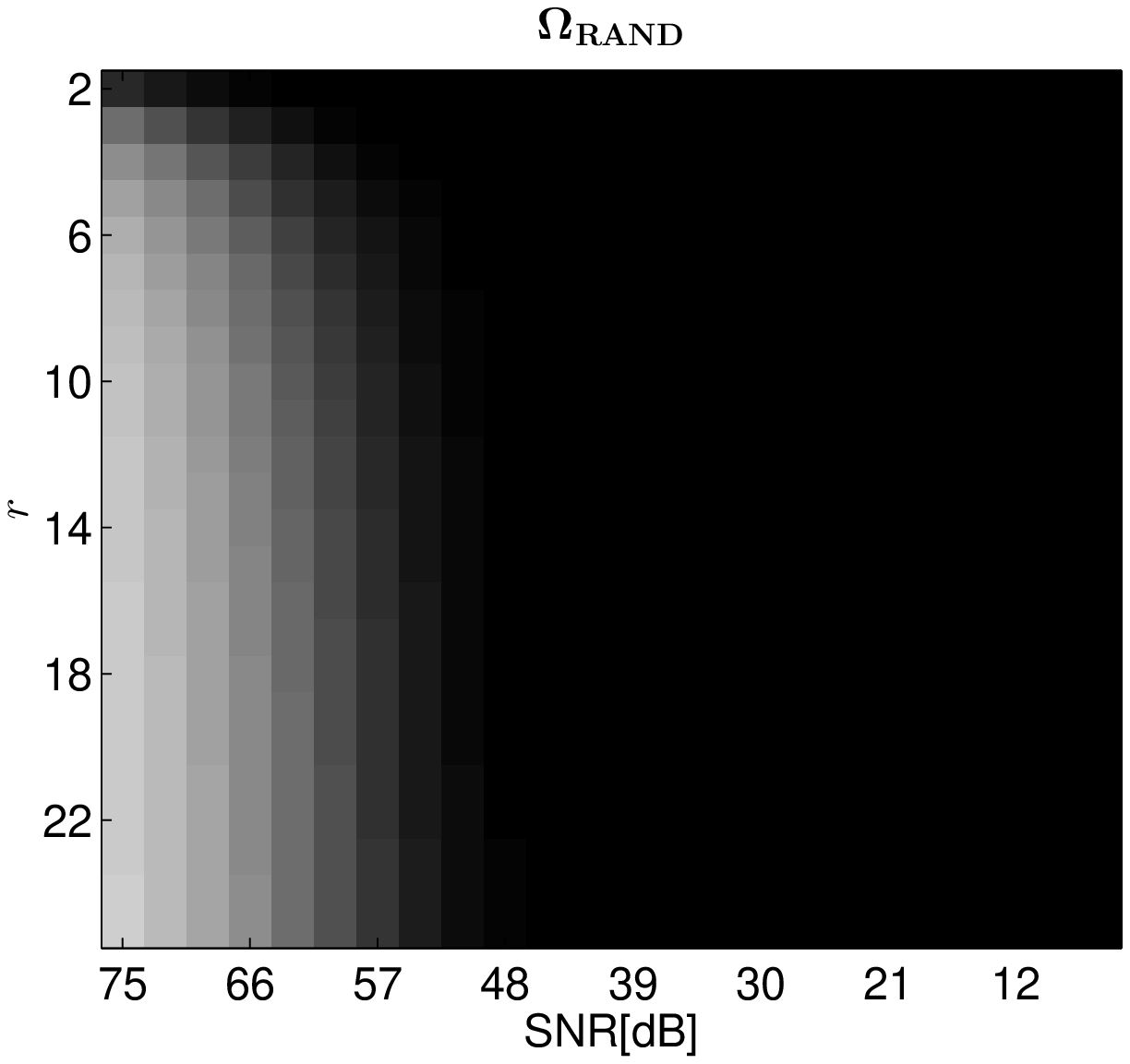}\\
\vspace{10pt}
\includegraphics[scale=0.4]{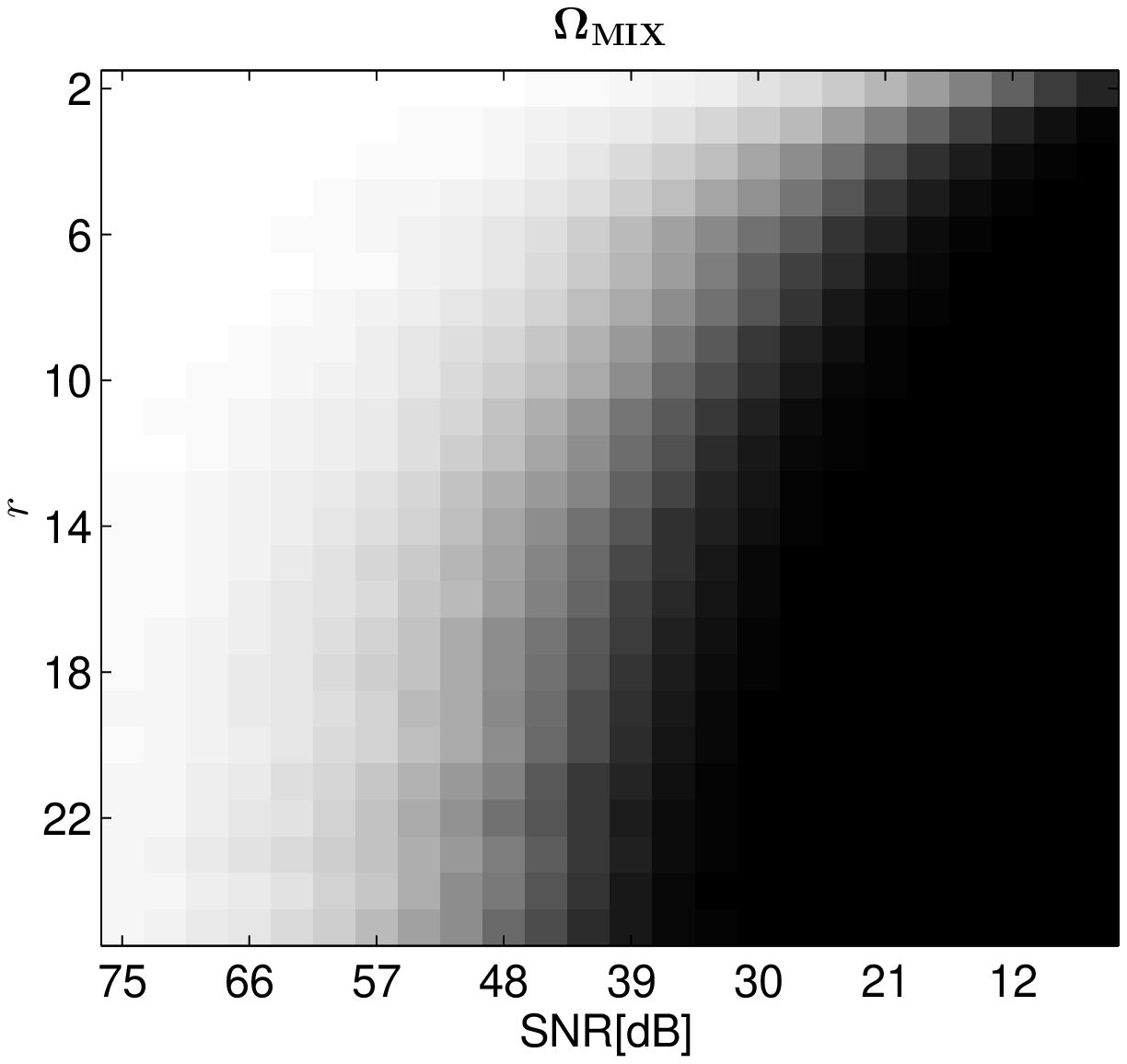} \hspace{20pt}
\includegraphics[scale=0.4]{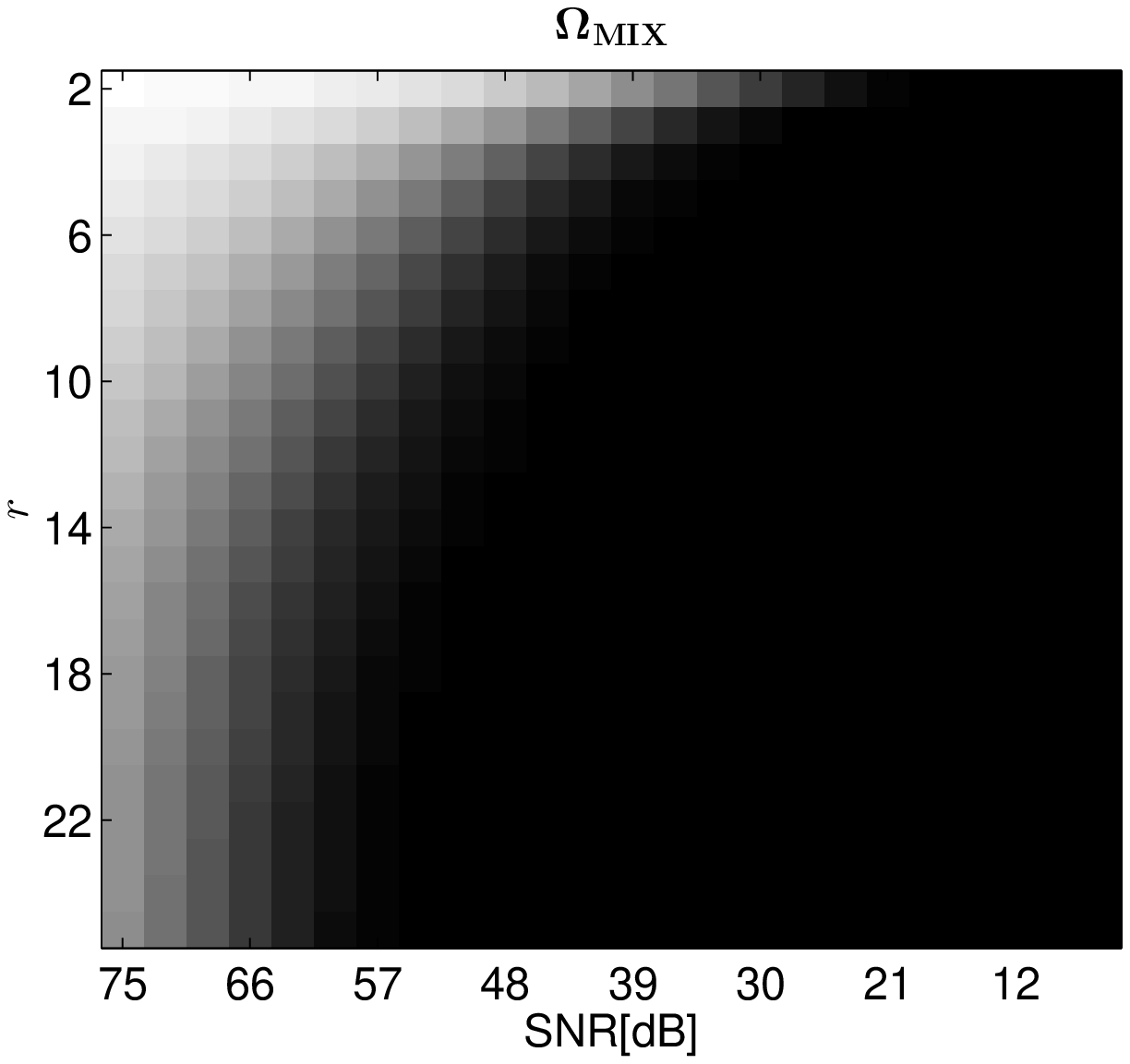}
\caption[]{Empirical ratios of success and their theoretical lower bounds for the thresholding algorithm on three types of analysis dictionaries of size $200\times 100$ for varying analysis subspace dimensions $r$ and SNR levels. For each pair of $r$ and SNR we generate $N=1000$ signals. The theoretical bounds are computed using Eq. (\ref{eq:TighterLowerBoundPrSuccess}) by plugging into it the empirical distribution of $\ell$ and $\alpha_r^\Lambda$, which is computed from these signals. Left: The empirical ratios of success. Right: The theoretical bounds.}
\label{fig:SuccessRatioHighDim}
\end{figure*}

Before concluding this section, we bring several additional experiments, this time with higher dimensional signals, in order to demonstrate the behavior of the thresholding algorithm, and the comparison between empirical performance and the theoretical forecasts. We consider signals of dimension $d=100$ and three types of analysis dictionaries (same as before), each with $p=200$ atoms. We test denoising setups where the true analysis subspace dimension $r$ varies in the range $[2,25]$ and the SNR in the range $6dB$ to $75dB$. For each pair of $r$ and noise level $\sigma$ we generate $N=1000$ signals. When evaluating the theoretical bounds, we cannot use the value of $\alpha_r$ as exhaustive search for its value is unfeasible. We therefore use the expression given in Eq. (\ref{eq:TighterLowerBoundPrSuccess}), where we plug into it an empirical distribution of the values of $\ell$ and $\alpha_r^\Lambda$ that is computed from the signal examples, instead of the exact one we have used for the low dimensional setups. The empirical ratios of success and their theoretical lower bounds are shown in Fig. \ref{fig:SuccessRatioHighDim} for the three types of analysis dictionaries of size $200$-by-$100$. Each of these ratios is displayed as a matrix where white corresponds to one and black corresponds to zero.

Several observations can be made from Fig. \ref{fig:SuccessRatioHighDim}. First, the general behavior of the three dictionary types remain as before: The performance is best for $\mOm_{DIF}$, second best for $\mOm_{MIX}$ and the worse for $\mOm_{RAND}$, both in terms of the empirical and the theoretical success rates. Secondly, for $\mOm_{DIF}$ and $\mOm_{MIX}$ the best performance is obtained for low SNR levels and low subspace dimensions $r$ (the top left corner of the matrix). This is a desired behavior due to the fact that we typically want a low subspace dimension, which improves the denoising performance. For $\mOm_{RAND}$ however, the best theoretical results are obtained for low SNR levels and high values of $r$ (the bottom left corner). The theoretical predictions for this dictionary are less reliable, as we can see that the actual performance is quite similar for all values of $r$.

\section{Relation to Existing Results}

\label{sec:Relations}

There are several exiting contributions in the published literature on developing pursuit algorithms for the co-sparse analysis model and studying their performance from a theoretical stand-point. Here we mention several papers that are of relevance to this work. We provide a brief review of their content, followed by a discussion on the relation to our results.

The first work we briefly refer to is \cite{AnalysisKSVD11}, which concentrates on the analysis dictionary learning problem. Two greedy analysis pursuit algorithms are developed for the denoising problem, as part of the overall learning paradigm -- these algorithms are the Backward Greedy (BG) and the Optimized BG (OBG). Both these algorithms are constrcuted by imitating synthesis based pursuit methods, and brought without a theoretical justification of any sort. Interestingly, the work in \cite{AnalysisKSVD11} provides an empirical evidence for the positive effect that strong linear dependencies within the analysis dictionary have on the success of pursuit algorithms. 

The work of \cite{Nam2011ICASSP,Nam2011Journal} considers a noise-free measurement setup where the co-sparse analysis signal is measured by $\vy=\mM\vx$, from which we would like to recover $\vx$. The authors of \cite{Nam2011ICASSP,Nam2011Journal} explore various uniqueness properties of this problem setup and suggest using either an analysis $\ell_1$-norm minimization or a Greedy-Analysis-Pursuit (GAP) algorithm (note that GAP is different from the above mentioned BG and OBG - see more in \cite{AnalysisKSVD11}) for recovering the signal. They analyze the performance of these pursuit algorithms for the noise-free setup, deriving a sufficient condition for success of both algorithms in terms of the analysis dictionary $\mOm$, the true co-support $\Lambda$ of $\vx$ and the null-space of $\mM$. Due to its apparent similarity to the ERC for the synthesis model, the derived condition is termed {\em analysis ERC}. 

The theoretical study of analysis $\ell_1$-norm based pursuit in a measurement setup is also the main focus of another recent work \cite{PeyreAnalysis2011}. This includes the derivation of conditions for noiseless identifiability and robustness to bounded noise, in terms of the sign pattern of $\mOm\vx$ and assuming that the null spaces of the measurement matrix $\mM$ and the analysis dictionary $\mOm$ intersect only at the zero vector. Note that all of the resulting conditions in \cite{Nam2011ICASSP,Nam2011Journal,PeyreAnalysis2011} are somewhat implicit, especially in the latter work, where the condition involves an inner optimization stage for a given sign pattern. This makes the derived conditions hard to interpret.

A different work altogether is proposed in \cite{Candes2011}. The authors \cite{Candes2011} suggest a hybrid viewpoint to the synthesis and analysis models, where the signal of interest is a synthesis-and-analysis signal, constructed as $\vx=\mD\va$ with a sparse synthesis representation $\va$. However, this signal is also characterized as an analysis signal in the sense that it has a small $\ell_1$ energy in the tail of the analysis representation $\mD^T\va$. They suggest using an analysis-based approach for recovering the signal from its undersampled and noisy measurements $\vy=\mM\vx+\ve$. Their approach is based on $\ell_1$-norm sparsity of $\mD^T\vx$ deriving a theoretical upper bound on the denoising error obtained by $\ell_1$ analysis pursuit in this setup. To obtain the desired bound they require the measurement matrix $\mM$ to satisfy a certain property adapted to $\mD$, termed D-RIP, which is similar to the well-known RIP aside from a delicate modification -- instead of bounding the $\ell_2$ norm of $\mM\vv$ for all $k$-sparse vectors $\vv$, the norm of $\mM\vv$ is bounded for all vectors $\vv$ that can be expressed as a linear combination of $k$ columns of $\mD$.

The work of \cite{Raja2011} suggests a family of new pursuit algorithms for recovering co-sparse analysis signals from their undersampled measurements. These algorithms are analogous to the synthesis-based iterative hard thresholding algorithm, with a modification of the projection step intended for adapting this framework to the analysis model. The authors of \cite{Raja2011} present theoretical recovery guarantees for these analysis pursuit algorithms in the noiseless setup, assuming that the measurement matrix satisfies the $\mOm$-RIP (an analysis counterpart for the D-RIP of \cite{Candes2011}).  

In this paper we focus on a denoising setup, similar to \cite{AnalysisKSVD11} and assume no measurement matrix. Our focus is the most simple analysis pursuit algorithm -- the thresholding. This allows us to remove some of the ambiguities that are present in previous works, where the resulting theoretical conditions mix both the measurement matrix $\mM$ and the analysis dictionary $\mOm$; we focus on internal properties of $\mOm$ only. Indeed, our derived theoretical guarantees are expressed in terms of the noise level, the co-sparsity $\ell$ of the signal over $\mOm$ and internal properties of $\mOm$. Instead of using dictionary measures that mimic the synthesis counterpart model, as practiced in \cite{Nam2011Journal}, which uses analysis ERC, or \cite{Candes2011,Raja2011}, which use RIP-like properties, we suggest a novel measure, termed {\em Restricted Orthogonal Projection Property} (ROPP), which seems to be more relevant to analysis dictionaries. This property is much more explicit than the one arising from the theoretical analysis of \cite{PeyreAnalysis2011}. Our derived results are simple to interpret, and specifically we see that strong linear dependencies improve the pursuit algorithm's success rate.

\section{Conclusions}

\label{sec:Conclusion}

In this work we have made an initial attempt at addressing the question of what makes an analysis dictionary suitable for co-sparse estimation. We have concentrated on a denoising setup and considered the use of a thresholding algorithm for the corresponding analysis pursuit problem. Our experiments show that this simple algorithm can perform quite well for certain analysis dictionaries, while failing on others. To better understand this behavior we further explored the performance of this algorithm in the presence of white Gaussian random noise, developing theoretical guarantees for the ability of the algorithm to recover the true underlying co-support. This study reveals two significant properties of an analysis dictionary that are key in dictating whether the pursuit will succeed or fail: The degree of linear dependencies between rows of $\mOm$ and the level of independence between subsets of rows and other atoms, a property we termed ROPP. We have found that it is desired to have many linear dependencies, as they increase the co-sparsity level. Similarly, the ROPP constant should be as high as possible. Finally, we have shown how the developed theoretical guarantees can explain our empirical results and predict them quite well. This work gives rise to various open questions that will be the topics of future research. These include topics such as these:
\begin{enumerate}
\item While this work concentrated on the thresholding algorithm, a similar theoretical study should be given to other pursuit algorithms. Perhaps the quality measures we identified in this work could be of help in such study. 
\item This work defines the success of the pursuit algorithm by the complete identification of the co-support. However, this algorithm may perform rather well (in denoising terms) even in situations where only part of the support has been found. Extending this work to cover such cases would improve our prediction for the range of success of the thresholding algorithm. 
\item How could we incorporate the proposed quality measures for $\mOm$ directly into the dictionary learning process? By doing so we may design better analysis dictionaries, which will ultimately lead to performance improvement and make the analysis model and its learned dictionary suitable for a wide range of processing applications.
\end{enumerate}

\bibliography{references}
\bibliographystyle{IEEEbib}

\end{document}

%% file: defs.tex
\newcommand{\va}{\boldsymbol{\alpha}}

\newcommand{\vd}{\textbf{d}}
\newcommand{\ve}{\textbf{e}}

\newcommand{\vg}{\boldsymbol{\gamma}}

\newcommand{\vq}{\textbf{q}}

\newcommand{\vu}{\textbf{u}}
\newcommand{\vv}{\textbf{v}}
\newcommand{\vw}{\textbf{w}}
\newcommand{\vx}{\textbf{x}}
\newcommand{\vy}{\textbf{y}}

\newcommand{\mA}{\textbf{A}}
\newcommand{\mB}{\textbf{B}}

\newcommand{\mD}{\textbf{D}}

\newcommand{\mI}{\textbf{I}}

\newcommand{\mM}{\textbf{M}}

\newcommand{\mOm}{\boldsymbol{\Omega}}
\newcommand{\mP}{\textbf{P}}

\def \RR {\mathbf{R}}
\newcommand{\zmin}{z_{min}}

\newtheorem{lemma}{Lemma}

\newtheorem{corollary}{Corollary}
\newtheorem{theorem}{Theorem}
\newtheorem{definition}{Definition}

\newcommand{\argminlong}[1]{{\hbox{$\underset{#1}{\text{Argmin}}\;$}}}

\renewcommand{\max}[1]{{\hbox{$\underset{#1}{\text{Max}}\;$}}}
\renewcommand{\min}[1]{{\hbox{$\underset{#1}{\text{Min}}\;$}}}

\newcommand{\st}{\text{Subject To}}

\newcommand{\Span}{\text{Span}}

\newcommand{\tr}{\text{tr}}
\newcommand{\rank}{\text{Rank}}
\newcommand{\E}{\text{E}}

\newcommand{\R}{\mathbb{R}}

\newcommand{\casebegin}{\left\{\begin{array}{ll}}
\newcommand{\caseend}{\end{array}\right.}

\newcommand{\ltriplebar}{\lvert\kern-0.9pt\lvert\kern-0.9pt\lvert}
\newcommand{\rtriplebar}{\rvert\kern-0.9pt\rvert\kern-0.9pt\rvert}

%% file: Analysis_THR_double.bbl
\begin{thebibliography}{10}

\bibitem{Review09}
A.M. Bruckstein, D.L. Donoho, and M.~Elad,
\newblock ``From sparse solutions of systems of equations to sparse modeling of
  signals and images,''
\newblock {\em SIAM Review}, vol. 51, no. 1, pp. 34--81, 2009.

\bibitem{MyBook}
M.~Elad,
\newblock {\em Sparse and Redundant Representations: From Theory to
  Applications in Signal and Image Processing},
\newblock Springer, 2010.

\bibitem{Donoho01}
D.L. Donoho and X.~Huo,
\newblock ``Uncertainty principles and ideal atomic decomposition,''
\newblock {\em {IEEE} Trans. on Inf. Theory}, vol. 47, no. 7, pp. 2845--2862,
  2001.

\bibitem{Donoho03}
D.~L. Donoho and M.~Elad,
\newblock ``Optimally sparse representation in general (nonorthogonal)
  dictionaries via l1 minimization,''
\newblock {\em Proc. Nat. Aca. Sci.}, vol. 100, pp. 2197--2202, 2003.

\bibitem{Tropp04}
J.A. Tropp,
\newblock ``Greed is good: Algorithmic results for sparse approximation,''
\newblock {\em {IEEE} Trans. on Inf. Theory}, vol. 50, no. 10, pp. 2231--2242,
  2004.

\bibitem{Candes05}
E.J. Candes and T.~Tao,
\newblock ``Decoding by linear programming,''
\newblock {\em {IEEE} Trans. on Inf. Theory}, vol. 51, no. 12, pp. 4203--4215,
  2005.

\bibitem{Candes06}
E.J. Candes, J.K. Romberg, and T.~Tao,
\newblock ``Stable signal recovery from incomplete and inaccurate
  measurements,''
\newblock {\em Comm. Pure Appl. Math.}, vol. 59, no. 8, pp. 1207--1223, 2006.

\bibitem{Shtok08}
J.~Shtok and M.~Elad,
\newblock ``Analysis of the basis pursuit via the capacity sets,''
\newblock {\em J. Fourier Anal. App.}, vol. 14, no. 5--6, pp. 688--711, 2008.

\bibitem{Nemirovski11}
A.~Juditsky and A.~Nemirovski,
\newblock ``On verifiable sufficient conditions for sparse signal recovery via
  l1 minimization,''
\newblock {\em Mathematical Programming: Series A and B - Special Issue on
  Optimization and Machine learning}, vol. 127, no. 1, pp. 57--88, 2011.

\bibitem{Schnass07}
K.~Schnass and P.~Vandergheynst,
\newblock ``Average performance analysis for thresholding,''
\newblock {\em {IEEE} Signal Process. Lett.}, vol. 14, no. 11, pp. 828--831,
  2007.

\bibitem{Ben-Haim10}
Z.~Ben-Haim, Y.C. Eldar, and M.~Elad,
\newblock ``Coherence-based performance guarantees for estimating a sparse
  vector under random noise,''
\newblock {\em {IEEE} Trans. on Signal Processing}, vol. 58, no. 10, pp.
  5030--5043, 2010.

\bibitem{Elad07}
M.~Elad, P.~Milanfar, and R.~Rubinstein,
\newblock ``Analysis versus synthesis in signal priors,''
\newblock {\em Inverse Problems}, vol. 23, no. 3, pp. 947--968, 2007.

\bibitem{Candes2011}
E.J. Candes, Y.C. Eldar, D.~Needell, and P.~Randall,
\newblock ``Compressed sensing with coherent and redundant dictionaries,''
\newblock {\em Applied Computational Harmonic Analysis}, vol. 31, no. 1, pp.
  59--73, 2011.

\bibitem{Nam2011Sampta}
S.~Nam, M.E. Davies, M.~Elad, and R.~Gribonval,
\newblock ``Cosparse analysis modeling,''
\newblock in {\em Proceedings of {SAMPTA}}, 2011.

\bibitem{Peyre2011Sampta}
G.~Peyr\'{e} and J.~Fadili,
\newblock ``Learning analysis sparsity priors,''
\newblock in {\em Proceedings of {SAMPTA}}, 2011.

\bibitem{Nam2011ICASSP}
S.~Nam, M.~Davies, M.~Elad, and R.~Gribonval,
\newblock ``Cosaprse analysis modeling -- uniqueness and algorithms,''
\newblock in {\em Proceedings of {ICASSP}}, 2011, pp. 5804--5807.

\bibitem{Raja2011}
R.~Giryes, S.~Nam, R.~Gribonval, and M.E. Davies,
\newblock ``Iterative cosparse projection algorithms for the recovery of
  cosparse vectors,''
\newblock in {\em Proceedings of {EUSIPCO}}, 2011.

\bibitem{Ophir2011}
B.~Ophir, M.~Elad, N.~Bertin, and M.D. Plumbley,
\newblock ``Sequential minimal eigenvalues -- an approach to analysis
  dictionary learning,''
\newblock in {\em Proceedings of {EUSIPCO}}, 2011.

\bibitem{Davies2011}
R.~Gribonval M.~Yaghoobi, S.~Nam and M.E. Davies,
\newblock ``Analysis operator learning for overcomplete cosparse
  representations,''
\newblock in {\em Proceedings of {EUSIPCO}}, 2011.

\bibitem{Nam2011Journal}
S.~Nam, M.~Davies, M.~Elad, and R.~Gribonval,
\newblock ``The cosaprse analysis model and algorithms,''
\newblock {\em \textnormal{submitted to} Applied Computational Harmonic
  Analysis}.

\bibitem{PeyreAnalysis2011}
S.~Vaiter, G.~Peyr\'{e}, C.~Dossal, and J.~Fadili,
\newblock ``Robust sparse analysis regulatization,''
\newblock {\em \textnormal{Technical Report HAL-00627452}}.

\bibitem{AnalysisKSVD11}
R.~Rubinstein, T.Faktor, and M.~Elad,
\newblock ``Analysis {K-SVD}: A dictionary-learning algorithm for the analysis
  sparse model,''
\newblock {\em \textnormal{submitted to} {IEEE} Trans. on Signal Processing}.

\bibitem{Elad06}
M.~Elad,
\newblock ``Sparse representations are most likely to be the sparsest
  possible,''
\newblock {\em {EURASIP} Journal on Applied Signal Processing}, vol. 2006, pp.
  1--12, 2006.

\bibitem{BP98}
S.S. Chen, D.L. Donoho, and M.A. Saunders,
\newblock ``Atomic decomposition by basis pursuit,''
\newblock {\em {SIAM} Journal on Scientific Computing}, vol. 20, pp. 33--61,
  1998.

\bibitem{MP93}
S.~Mallat and Z.~Zhang,
\newblock ``Matching pursuits with time-frequency dictionaries,''
\newblock {\em {IEEE} Trans. Signal Processing}, vol. 41, pp. 3397--3415, 1999.

\bibitem{OMP93}
Y.C. Pati, R.~Rezaiifar, and P.S. Krishnaprasad,
\newblock ``Orthogonal matching pursuit: Recursive function approximation with
  applications to wavelet decomposition,''
\newblock in {\em Proceedings of the 27th Asilomar Conference on Signals,
  Systems and Computers}, 1993, vol.~1, pp. 40--44.

\bibitem{Sidak67}
Z.~\^{S}id\'{a}k,
\newblock ``Rectangular confidence regions for the means of multivariate normal
  distributions,''
\newblock {\em J. Amer. Statist. Assoc.}, vol. 62, no. 318, pp. 626--633, 1967.

\end{thebibliography}
